\documentclass[aps,preprint,preprintnumbers,nofootinbib]{revtex4}
\usepackage{mathtools}
\usepackage{makeidx}
\usepackage{eurosym}
\usepackage{graphicx}
\usepackage{bm}
\usepackage{fancyhdr}
\usepackage{amsmath}
\usepackage{amsfonts}
\usepackage{amsbsy}
\usepackage{amssymb}
\usepackage[mathscr]{eucal}
\usepackage{multirow}
\usepackage{mciteplus}
\usepackage{color}

\newcommand{\beq}{\begin{equation}}
\newcommand{\eeq}{\end{equation}}
\newcommand{\ba}{\begin{array}}
\newcommand{\ea}{\end{array}}
\newcommand{\bea}{\begin{eqnarray}}
\newcommand{\eea}{\end{eqnarray}}
\newcommand{\bseq}{\begin{subequations}}
\newcommand{\eseq}{\end{subequations}}

\begin{document}

\title{Uncertainty relations for time averaged weak values}
\author{Eli Pollak$^{1}$ and Salvador Miret-Art{\'{e}}s$^{2}$ }
\affiliation{$^1$ Chemical and Biological Physics Department, Weizmann Institute of
Science, 76100 Rehovoth, Israel}
\affiliation{$^2$ Instituto de F\'isica Fundamental, Consejo Superior de Investigaciones
Cient\'ificas, Serrano 123, 28006 Madrid, Spain}



\begin{abstract}
Time averaging of weak values using the quantum transition path time probability distribution enables us to establish a general
uncertainty principle for the weak values of two not necessarily Hermitian operators.  This new principle is a weak value analog of the Heisenberg-Robertson strong value uncertainty principle.  It leads to the conclusion that
it is possible to determine with high accuracy the simultaneous mean weak values of non-commuting operators by judicious choice of the pre- and
post-selected states. Generally, when the time fluctuations of the two weak values are proportional to each other there is no uncertainty limitation on their variances and, in principle, their means can be determined with arbitrary precision even though their corresponding operators do not commute. To exemplify these properties we consider specific weak value uncertainty relations for the time-energy, coordinate-momentum and coordinate-kinetic energy pairs. In addition we analyze spin operators and the Stern-Gerlach experiment in weak and strong inhomogeneous magnetic fields.  This classic case leads to anomalous spin values when the field is weak.   However, anomalous spin values are also associated with large variances implying that their measurement demands increased signal averaging.  These examples establish the importance of considering the time dependence of weak values in scattering experiments.

\end{abstract}

\maketitle
	



\renewcommand{\theequation}{1.\arabic{equation}} \setcounter{section}{0} %
\setcounter{equation}{0}

\section{Introduction}

\renewcommand{\theequation}{2.\arabic{equation}} \setcounter{section}{1} %
\setcounter{equation}{0}

The uncertainty principle is a fundament of quantum mechanics. The "intrinsic" uncertainty principle was derived by Robertson \cite{robertson} in 1929. He showed that the product of the standard deviations of the diagonal elements of two non-commuting Hermitian operators is bounded from below by the diagonal element of their commutator. For the coordinate momentum pair this lower  bound is $\hbar/2$. Heisenberg, in his consideration of an uncertainty principle seems to have had something else in mind, namely a measurement disturbance relation whereby measurement of one property of a system say its position creates an uncertainty in its momentum. It was only fifteen years ago that the fundamental measurement disturbance relation was derived by Ozawa \cite{ozawa03}, who also showed that this relation is not necessarily bounded from below by the Robertson lower bound.

The pioneering work by Aharonov et al. \cite{aharonov88} which showed among others that the results of weak measurement may be described by weak values of operators has attracted the attention of many theorists
\cite{wiseman07,nori12,hiley14} and experimentalists \cite{steinberg11,lundeen11,tamir13,dressel14}.  This led also to the consideration of the measurement disturbance relation in the context of weak values by Lund and Weisman \cite{lund10}. They showed how Ozawa's result could be measured experimentally by considering weak measurements and weak values. Their protocol was then implemented experimentally by Steinberg and coworkers. \cite{rozema12}

The uncertainty relation developed in this paper is somewhat different.  It is the application of the Robertson type of intrinsic uncertainty relation within the context of weak values and time evolution. The model underlying our considerations is a scattering experiment which for simplicity occurs in one dimension. A particle is scattered by a potential. It originates at the initial time to the left of the potential, interacts with it and is then either transmitted to its right where it continues motion as a free particle or reflected likewise to the left. A "screen" is set up to the right of the potential and the momentum of the particle is weakly measured prior but arbitrarily close to the location of the screen. In this way the momentum weak value is measured at the post selected location of the screen. The particle, if it is transmitted, will arrive at the screen at varying time and therefore with varying values of the weak momentum. This creates a distribution in time of momentum weak values. We are interested in the statistics of this distribution and this leads us to a weak value Robertson type uncertainty principle. It is intrinsic yet at the same time intimately related to the weak value.

The probability distribution which determines the time statistics is  the so-called
quantum transition path time probability distribution \cite{eli1}. In Ref. \cite{eli2} we showed how such time averaging of weak values leads to a well defined time-energy uncertainty principle and a time-energy weak value commutation relation. We also showed how with suitable post selection it is possible
to predict in advance the momentum of a single particle at a post-selected position with an accuracy greater than the Robertson limit of $\hbar / 2$. In these studies time is considered to be an external parameter as discussed by Aharonov and Bohm \cite{aharonov61} and Busch.
\cite{busch08}.

The topic of this paper is to consider the general scenario of the time statistics of weak values to derive a Robertson like uncertainty relation for the time averaged standard deviation of non-commuting, not necessarily Hermitian operators. To implement this goal we develop in Section II an algebra of what we define to be post-selected operators, their commutation relations and time averages. This is then used to derive a weak value uncertainty principle for the product of the standard deviations of the post-selected operators in terms of their time averaged commutators and anti-commutators.

We then consider in Section III some specific cases which exemplify the utility of the weak value uncertainty relation. The first example is the re-derivation of the time energy uncertainty relation for weak values. We then consider time dependent weak values for post-selected (normalized) coherent states. Due to this choice the time averaged weak values of the coordinate and momentum are found to be linearly proportional to each other. This implies through the weak value uncertainty relation that by judicious choice of the pre and post selected states both quantities can be determined with arbitrary precision. We then show how any uncertainty in the time averaged weak value of the position coordinate leads to an uncertainty in the time averaged weak value of the kinetic energy. Through the time energy uncertainty relation this means that increasing the accuracy in the determination of the position can lead to an increased indeterminacy in the time at which the weak value can be measured.

As a fourth example we choose the paradigmatic Stern-Gerlach experiment which is analyzed in terms of time dependent spin weak values in the presence of weak and strong inhomogeneous magnetic fields. Here too the fluctuations of the weak values of the spin about their mean also display
a proportionality so that the uncertainty principle does not impose any restriction on the time averaged spin weak values even though the
corresponding operators do not commute. Interestingly, the weak value uncertainty relation leads to the conclusion that the magnitude of the standard deviation of the spin weak value  is proportional to the magnitude of the time weighted mean spin weak value. This means that anomalously large spin weak values lead to a large uncertainty, implying, from an experimental point of view, that the weak values have to be measured many times
in order to obtain a reliable estimate of the mean value. This sheds light on recent neutron interferometry experiments \cite{denkmayr17,sponar18} where it was found that strong interaction measurement outperforms weak interaction when it comes to precision and accuracy of the measurement.

To further elucidate the implications of the weak value uncertainty relation we consider in Section IV two specific models. The first is one dimensional potential scattering to exemplify the coordinate, momentum, energy and time uncertainties. The second is a numerical analysis of the Stern-Gerlach setup in the presence of weak and strong inhomogeneous magnetic fields. Perhaps the most striking result here is that anomalous weak values are associated with large standard deviations as compared to the "regular" ones. Finally,
in Section V, we further discuss the implications of the weak value uncertainty principle and the importance of taking into consideration the time evolution of weak values.

\section{Time averaging and a weak value uncertainty principle}

\subsection{Weak value operator algebra}

In this Section we consider and develop an operator formalism for weak values. We
assume a post-selected normalized state $|\Phi \rangle $ and define the
density operator associated with it as
\begin{equation}
\hat{D}_{\Phi }=|\Phi \rangle \langle \Phi |.  \label{2.1}
\end{equation}%
The density operator is by definition Hermitian and it is a projection
operator, that is $\hat{D}_{\Phi }^{2}=\hat{D}_{\Phi }$. The weak value
operator $\hat{O}_{\Phi }$ associated with the operator $\hat{O}$ and the
post-selected state is defined as the product
\begin{equation}
\hat{O}_{\Phi }=\hat{D}_{\Phi }\hat{O}.  \label{2.2}
\end{equation}%
With this definition the weak value of
the operator $\hat{O}$ associated with the pre-selected (normalized)\ stated
$|\Psi \rangle $ and post-selected state $|\Phi \rangle $ is given by a ratio of strong values \cite{cohen18} with the pre-selected state:
\begin{equation}
O_{w}\left( \Phi ;\Psi \right) =\frac{\left\langle \Phi \left\vert \hat{O}%
\right\vert \Psi \right\rangle }{\left\langle \Phi |\Psi \right\rangle }=%
\frac{\left\langle \Psi \left\vert \hat{O}_{\Phi }\right\vert \Psi
\right\rangle }{\left\langle \Psi \left\vert \hat{D}_{\Phi }\right\vert \Psi
\right\rangle }.  \label{2.3}
\end{equation}%

A central goal of this paper is to present time averaging of weak values with corresponding general uncertainty relations. For
the chosen pre- and post-selected states the time probability distribution is
defined to be \cite{eli1}%
\begin{equation}
P_{\Psi }\left( t;\Phi \right) =\frac{\left\langle \Psi _{t}\left\vert \hat{D%
}_{\Phi }\right\vert \Psi _{t}\right\rangle }{\int_{0}^{\infty
}dt\left\langle \Psi _{t}\left\vert \hat{D}_{\Phi }\right\vert \Psi
_{t}\right\rangle }\equiv \frac{\left\langle \Psi _{t}\left\vert \hat{D}%
_{\Phi }\right\vert \Psi _{t}\right\rangle }{N\left( \Phi ;\Psi \right) }
\label{2.4}
\end{equation}%
where $|\Psi _{t}\rangle $ is the time evolved pre-selected state under the
Hamiltonian $\hat{H}$ :%
\begin{equation}
|\Psi _{t}\rangle =\exp \left( -\frac{i\hat{H}t}{\hbar }\right) |\Psi
\rangle .  \label{2.5}
\end{equation}%
The time averaged weak value is then calculated as %
\begin{eqnarray}
\left\langle O_{w}\left( \Phi ;\Psi \right) \right\rangle &\equiv
&\int_{0}^{\infty }dtP_{\Psi }\left( t;\Phi \right) O_{w}\left( \Phi ;\Psi
_{t}\right)  \notag \\
&=&\frac{1}{N\left( \Phi ;\Psi \right) }\int_{0}^{\infty }dt\left\langle
\Psi _{t}\left\vert \hat{O}_{\Phi }\right\vert \Psi _{t}\right\rangle
\label{2.6}
\end{eqnarray}%
thus relating time averaged weak values to time integrals of diagonal matrix
elements of weak valued operators.

We then consider two not necessarily Hermitian operators $\hat{A%
}$ and $\hat{B}$\ such that their commutation relation is the anti-Hermitian
operator%
\begin{equation}
\left[ \hat{B}^{\dag },\hat{A}\right] \equiv \hat{B}^{\dag }\hat{A}-\hat{A}%
^{\dag }\hat{B}  \label{2.7}
\end{equation}%
and their anti-commutation relation is the Hermitian operator
\begin{equation}
\left\{ \hat{B}^{\dag },\hat{A}\right\} \equiv \hat{B}^{\dag }\hat{A}+\hat{A}%
^{\dag }\hat{B}.  \label{2.8}
\end{equation}%
The commutation and anti-commutation relations for the
associated weak value operators are then:%
\begin{equation}
\left[ \hat{B}_{\Phi }^{\dag },\hat{A}_{\Phi }\right] \equiv \hat{B}^{\dag }%
\hat{D}_{\Phi }\hat{A}-\hat{A}^{\dag }\hat{D}_{\Phi }\hat{B}  \label{2.9}
\end{equation}%
\begin{equation}
\left\{ \hat{B}_{\Phi }^{\dag },\hat{A}_{\Phi }\right\} \equiv \hat{B}^{\dag
}\hat{D}_{\Phi }\hat{A}+\hat{A}^{\dag }\hat{D}_{\Phi }\hat{B}  \label{2.10}
\end{equation}%
and one notes that they are not the same as the original 
relations. 

To see this in more detail let us consider the specific example
in which the operators $\hat{A}$ and $\hat{B}$\ are Hermitian and the
post-selected state is an eigenfunction of the operator $\hat{B}$:%
\begin{equation}
\hat{B}|\Phi _{B}\rangle =\beta |\Phi _{B}\rangle .  \label{2.11}
\end{equation}%
One then has that%
\begin{equation}
\left[ \hat{B}_{\Phi _{B}},\hat{A}_{\Phi _{B}}\right] =\beta \left[ \hat{D}%
_{\Phi _{B}},\hat{A}\right] .  \label{2.12}
\end{equation}%
To be more concrete, we choose as an example the coordinate $\hat{q}$ and momentum $%
\hat{p}$ operators for which the commutator is $i\hbar $ so that
the\ off diagonal coordinate matrix element of the commutator is%
\begin{equation}
\left\langle x\left\vert \left[ \hat{q},\hat{p}\right] \right\vert x^{\prime
}\right\rangle =i\hbar \delta \left( x-x^{\prime }\right) .  \label{2.13}
\end{equation}%
However the off diagonal coordinate matrix element for the weak value
commutator is quite different%
\begin{equation}
\left\langle x\left\vert \left[ \hat{q}_{\Phi }^{\dag },\hat{p}_{\Phi }%
\right] \right\vert x^{\prime }\right\rangle =i\hbar \left[ x\left\langle
x|\Phi \right\rangle \frac{\partial \left\langle \Phi |x^{\prime
}\right\rangle }{\partial x^{\prime }}+\frac{\partial \left\langle x|\Phi
\right\rangle }{\partial x}x^{\prime }\left\langle \Phi |x^{\prime
}\right\rangle \right] .  \label{2.14}
\end{equation}%
This has of course implications when considering as below the weak value
uncertainty relations.

\subsection{Weak value uncertainty relations}

The deviation of the post-selected operator $\hat{O}_{\Phi }$ from its time
averaged mean is denoted as:%
\begin{equation}
\Delta \hat{O}_{\Phi }=\hat{O}_{\Phi }-|\Phi \rangle \langle \Phi
|\left\langle O_{w}\left( \Phi ;\Psi \right) \right\rangle .  \label{2.15}
\end{equation}%
We then consider as before two operators, not necessarily Hermitian, $\hat{A}
$ and $\hat{B}$. Following the Robertson derivation of the Heisenberg
uncertainty principle \cite{robertson,cohen-tannoudji} one constructs the following inequality
\begin{equation}
0\leq \frac{1}{N\left( \Phi ;\Psi \right) }\int_{0}^{\infty }dt\langle \Psi
_{t}|\left[ \Delta \hat{A}_{\Phi }^{\dag }+i\lambda ^{\ast }\Delta \hat{B}%
_{\Phi }^{\dag }\right] \left[ \Delta \hat{A}_{\Phi }-i\lambda \Delta \hat{B}%
_{\Phi }\right] |\Psi _{t}\rangle  \label{2.16}
\end{equation}%
where $\lambda $ is an arbitrary complex number%
\begin{equation}
\lambda =\lambda _{R}+i\lambda _{I}.  \label{2.17}
\end{equation}

As we shall see below when analyzing some specific examples, it may happen
that even if the two operators $\hat{A}$ and $\hat{B}$ do not commute
their weak values are proportional to each other, that is,  $\left\langle
\Phi \left\vert \Delta \hat{A}_{\Phi }\right\vert \Psi _{t}\right\rangle
=C\left\langle \Phi \left\vert \Delta \hat{B}_{\Phi }\right\vert \Psi
_{t}\right\rangle $ with $C$ a proportionality constant. In such a case,
the value $i\lambda =C$ causes the r.h.s of Eq. (\ref{2.16}) to vanish and there is no constraint on the variances so both may vanish. In other words, time averaged weak
values of non-commuting operators may at times be obtained with certainty.

In the general case, when such a linear relation does not occur,
minimizing the right hand side of the inequality in Eq. (\ref{2.16}) with
respect to the two time independent real numbers $\lambda _{R}$ and $\lambda
_{I}$ leads to the time averaged weak value uncertainty relation
\begin{eqnarray}
&&\frac{\int_{0}^{\infty }dt\langle \Psi _{t}|\Delta \hat{A}_{\Phi }^{\dag
}\Delta \hat{A}_{\Phi }|\Psi _{t}\rangle }{N\left( \Phi ;\Psi \right) }\frac{%
\int_{0}^{\infty }dt\langle \Psi _{t}|\Delta \hat{B}_{\Phi }^{\dag }\Delta
\hat{B}_{\Phi }|\Psi _{t}\rangle }{N\left( \Phi ;\Psi \right) }\geq  \notag
\\
&&\cdot\frac{1}{4}\left( \left[ \frac{\int_{0}^{\infty }dt\langle \Psi
_{t}|\left\{ \Delta \hat{B}_{\Phi }^{\dag },\Delta \hat{A}_{\Phi }\right\}
|\Psi _{t}\rangle }{N\left( \Phi ;\Psi \right) }\right] ^{2}-\left[ \frac{%
\int_{0}^{\infty }dt\langle \Psi _{t}|\left[ \Delta \hat{B}_{\Phi }^{\dag
},\Delta \hat{A}_{\Phi }\right] |\Psi _{t}\rangle }{N\left( \Phi ;\Psi
\right) }\right] ^{2}\right) .  \label{2.18}
\end{eqnarray}%
which is the central formal result of this paper. It expresses the minimal
standard deviation of the time average of the weak values of two operators
whose weak value commutators and anti-commutators do not vanish. This weak
value uncertainty principle is a direct analog of the strong value
uncertainty principle. However, due to the post-selection, in practice, it
differs from it significantly. This formal result may be rewritten also as:%
\begin{eqnarray}
&&\left\langle \left\vert \Delta A_{w}\left( \Phi ;\Psi _{t}\right)
\right\vert ^{2}\right\rangle \left\langle \left\vert \Delta B_{w}\left(
\Phi ;\Psi _{t}\right) \right\vert ^{2}\right\rangle \geq  \notag \\
&&\frac{1}{4}\left( \left\langle \Delta B_{w}^{\ast }\left( \Phi ;\Psi
_{t}\right) \Delta A_{w}\left( \Phi ;\Psi _{t}\right) +\Delta B_{w}\left(
\Phi ;\Psi _{t}\right) \Delta A_{w}^{\ast }\left( \Phi ;\Psi _{t}\right)
\right\rangle \right) ^{2}  \notag \\
&&-\frac{1}{4}\left( \left\langle \Delta B_{w}^{\ast }\left( \Phi ;\Psi
_{t}\right) \Delta A_{w}\left( \Phi ;\Psi _{t}\right) -\Delta B_{w}\left(
\Phi ;\Psi _{t}\right) \Delta A_{w}^{\ast }\left( \Phi ;\Psi _{t}\right)
\right\rangle \right) ^{2}  \notag \\
&=&\left\vert \left\langle \Delta A_{w}\left( \Phi ;\Psi _{t}\right) \Delta
B_{w}^{\ast }\left( \Phi ;\Psi _{t}\right) \right\rangle \right\vert ^{2}
\label{2.19}
\end{eqnarray}%
where we have used the notation for the time dependent fluctuation of the weak values%
\begin{equation}
\Delta A_{w}\left( \Phi ;\Psi _{t}\right) =A_{w}\left( \Phi ;\Psi
_{t}\right) -\left\langle A_{w}\left( \Phi ;\Psi _{t}\right) \right\rangle .
\label{2.20}
\end{equation}%
This result is somewhat more general than the formal uncertainty expression
given in Eq. (\ref{2.18}) since it is valid also in the case that the post-selected state is not normalizable, for example it might be the continuum
state $|x\rangle $. \qquad

From Eq. (\ref{2.19}), we note that if the post-selected state is an
eigenfunction of either the operator $\hat{A}$ or $\hat{B}$ then there is no
uncertainty. Without loss of generality, let us suppose that the
post-selected state is an eigenfunction of $\hat{A}$ denoted as $|a\rangle $
with complex eigenvalue $a$, then one sees that the weak value of $%
\hat{A}$ is
\begin{equation}
A_{w}\left( a;\Psi _{t}\right) =\frac{\left\langle a\left\vert \hat{A}%
\right\vert \Psi _{t}\right\rangle }{\left\langle a|\Psi _{t}\right\rangle }%
=a^{\ast }  \label{2.21}
\end{equation}%
and it is independent of the time. This means that the
time average of the weak value is the same as the eigenvalue so that there
is no time dependent fluctuation in the weak value and $\Delta A_{w}\left( \Phi
;\Psi \right) =0$. 

An important example is when the post-selected
state is $|x\rangle $ and the two operators are the coordinate and momentum
operators. There is no uncertainty in determining precisely the weak values
of the coordinate and momentum simultaneously, that is, at the same elapsed
time. This reflects a fundamental difference between strong and weak values.
In the case of strong values, if the operators do not commute, the
uncertainty relation is never trivial. For weak values, even if the
operators do not commute, the measure of uncertainty depends on the choice
of the post-selected state and may vanish. It is thus possible to know with
high accuracy the mean weak values of non-commuting operators
by judicious choice of the pre- and post-selected states. A specific case,
exemplifying this result, has been previously considered \cite{eli2}  where we showed that
in a scattering experiment, using a pre-selected coherent state, one
may know after the event is over the simultaneous values of the coordinate
and the momentum.

\renewcommand{\theequation}{3.\arabic{equation}} \setcounter{section}{2} %
\setcounter{equation}{0}

\section{Examples of weak value uncertainty relations}

\subsection{The weak value time-energy uncertainty relation}

The weak value time-energy uncertainty relation discussed in Ref. \cite{eli2} is considered here in the framework of the general formalism of the previous section. The
definition of a time operator is even today unsolved, in the sense that
there is no Hermitian time operator $\hat{T}$ whose commutation relation
with the energy operator $\hat{H}$ is $i\hbar $. Instead of employing a time operator
we consider it as a parameter
multiplying the identity operator $\hat{I}$. Its weak value for any pre- and
post-selected states is therefore
\begin{equation}
t_{w}\left( \Phi ;\Psi _{t}\right) =\frac{\left\langle \Phi \left\vert t\hat{%
I}\right\vert \Psi _{t}\right\rangle }{\left\langle \Phi |\Psi
_{t}\right\rangle }=t  ,  \label{3.1}
\end{equation}%
and its time average
\begin{equation}
\left\langle t_{w}\left( \Phi ;\Psi \right) \right\rangle =\int_{0}^{\infty
}dtP_{\Psi }\left( t;\Phi \right) t  \label{3.2}
\end{equation}%
will depend on the pre- and post-selected states.

The energy weak value is by definition:%
\begin{equation}
E_{w}\left( \Phi ;\Psi _{t}\right) =\frac{\left\langle \Phi \left\vert \hat{H%
}\right\vert \Psi _{t}\right\rangle }{\left\langle \Phi |\Psi
_{t}\right\rangle }=i\hbar \frac{\partial }{\partial t}\ln \left\langle \Phi
|\Psi _{t}\right\rangle  \label{3.3}
\end{equation}%
and its mean may be written as%
\begin{equation}
\left\langle E_{w}\left( \Phi ;\Psi _{t}\right) \right\rangle =\frac{i\hbar
}{N\left( \Phi ;\Psi \right) }\int_{0}^{\infty }dt\left\langle \Psi
_{t}|\Phi \right\rangle \frac{\partial }{\partial t}\left\langle \Phi |\Psi
_{t}\right\rangle  \label{3.4}
\end{equation}%
from which it becomes clear that
\begin{equation}
\text{Im}\left\langle E_{w}\left( \Phi ;\Psi _{t}\right) \right\rangle =-%
\frac{\hbar }{2N(\Phi ;\Psi )}|\langle \Phi |\Psi _{0}\rangle |^{2}.
\label{3.5}
\end{equation}%
Initially, the pre- and post-selected states typically do not overlap so
that the imaginary energy value vanishes. We then find, using an integration
by parts, that the time mean of the time-energy weak value commutator is given by:%
\begin{eqnarray}
&&\left\langle \left[ \Delta t_{w}\left( \Phi ;\Psi _{t}\right) ,\Delta
E_{w}\left( \Phi ;\Psi _{t}\right) \right] \right\rangle  \notag \\
&=&\left\langle \left( t-\left\langle t_{w}\left( \Phi ;\Psi \right)
\right\rangle \right) \left[ \Delta E_{w}^{\ast }\left( \Phi ;\Psi _{t}\right)
-\Delta E_{w}\left( \Phi ;\Psi _{t}\right) \right] \right\rangle
\notag \\
&=&i\hbar -i\hbar \langle t_{w}(\Phi ;\Psi )\rangle \frac{|\langle \Phi
|\Psi _{0}\rangle |^{2}}{N(\Phi ;\Psi )}.  \label{3.6}
\end{eqnarray}
In contrast to the difficulty in defining time operators which obey such a
commutation relation, we find that when using time averaging of weak values,
the energy-time commutation relation emerges naturally. Furthermore, when
the initial overlap of the pre- and post-selected states vanishes one regains
the "standard" ($i\hbar $) value for the commutator. It is not possible to
obtain such a general result for the anti-commutator, for which one needs
more knowledge about the specific system and pre- and post-selection under
study.

The result for the commutator then implies the time-energy weak value
uncertainty relation%
\begin{eqnarray}
\left\langle \Delta t_{w}\left( \Phi ;\Psi \right) \Delta t_{w}^{\ast }\left( \Phi
;\Psi \right) \right\rangle \left\langle \Delta E_{w}\left( \Phi
;\Psi \right) \Delta E_{w}^{\ast }\left( \Phi ;\Psi \right) \right\rangle
&\geq &\frac{\hbar ^{2}}{4}\left( 1-\langle t_{w}(\Phi ;\Psi )\rangle \frac{%
|\langle \Phi |\Psi _{0}\rangle |^{2}}{N(\Phi ;\Psi )}\right) ^{2}  \notag \\
&&  \label{3.7}
\end{eqnarray}%
and this reduces to the "standard" estimate of $\hbar ^{2}/4$ when initially
the overlap of the pre- and post-selected states vanishes.

\subsection{The coordinate-momentum weak value uncertainty with post-selected coherent states}

To provide some practical results, we will study the weak values of the
coordinate and momentum assuming that the post-selected state is the
coherent state%
\begin{equation}
\left\langle x|\Phi \right\rangle =\left( \frac{\Gamma }{\pi }\right)
^{1/4}\exp \left( -\frac{\Gamma }{2}\left( x-x_{f}\right) ^{2}+\frac{i}{%
\hbar }p_{f}\left( x-x_{f}\right) \right) .  \label{3.8}
\end{equation}%
The coordinate and momentum weak values at time $t$ are then%
\begin{equation}
x_{w}\left( \Phi ;\Psi _{t}\right) =\frac{\left\langle \Phi \left\vert
x\right\vert \Psi _{t}\right\rangle }{\left\langle \Phi |\Psi
_{t}\right\rangle },\text{ \ \ \ \ }p_{w}\left( \Phi ;\Psi _{t}\right) =%
\frac{\left\langle \Phi \left\vert p\right\vert \Psi _{t}\right\rangle }{%
\left\langle \Phi |\Psi _{t}\right\rangle }.  \label{3.9}
\end{equation}%
Due to the Gaussian form of the post-selected coherent state we find that
the weak values of the momentum and the coordinate are linearly related
\begin{equation}
p_{w}\left( \Phi ;\Psi _{t}\right) =p_{f}+i\hbar \Gamma \left[
x_{f}-x_{w}\left( \Phi ;\Psi _{t}\right) \right] .  \label{3.10}
\end{equation}%
This implies a linear connection between the time averaged weak value
of the momentum and the coordinate:%
\begin{equation}
\left\langle p_{w}\left( \Phi ;\Psi _{t}\right) \right\rangle =p_{f}+i\hbar
\Gamma \left[ x_{f}-\left\langle x_{w}\left( \Phi ;\Psi _{t}\right)
\right\rangle \right]  \label{3.11}
\end{equation}%
so that%
\begin{equation}
\Delta p_{w}\left( \Phi ;\Psi _{t}\right) =-i\hbar \Gamma \Delta x_{w}\left(
\Phi ;\Psi _{t}\right) .  \label{3.12}
\end{equation}%
In other words, the time induced fluctuations in the weak values of the coordinate and
the momentum are proportional to each other. As noted in the previous
Section, such a linear relation implies that there is no uncertainty
limitation on the variances of the time averaged weak values of the momentum
and coordinate and, in principle, they can be known simultaneously with
arbitrary precision even though the operators themselves do not commute.

\subsection{The coordinate-kinetic energy weak value uncertainty with post-selected coherent states}

In contrast to the coordinate-momentum pair, when one considers the kinetic
energy-coordinate pair, the corresponding time averaged weak value uncertainty relation
sets a limit on the accuracy with which one may simultaneously determine
both weak values. As before, we use a post-selected coherent state Eq. (\ref{3.8}). One then readily finds that
\begin{eqnarray}
T_{w}\left( \Phi ;\Psi _{t}\right)  &=&\frac{\int_{-\infty }^{\infty
}dx\left\langle \Phi |p^{2}\left\vert x\rangle \langle x\right\vert \Psi
_{t}\right\rangle }{2M\left\langle \Phi |\Psi _{t}\right\rangle }  \notag \\
&=&\frac{\hbar ^{2}\Gamma }{2M}+\frac{\left( p_{f}+i\hbar \Gamma
x_{f}\right) ^{2}}{2M}-\frac{i\hbar \Gamma \left[ p_{f}+i\hbar \Gamma x_{f}%
\right] }{M}x_{w}\left( \Phi ;\Psi _{t}\right) -\frac{\hbar ^{2}\Gamma ^{2}}{%
2M}x_{w}^{2}\left( \Phi ;\Psi _{t}\right)   \notag \\
&&  \label{3.13}
\end{eqnarray}%
so that the fluctuation of the weak value of the kinetic energy operator is
related to the fluctuations of the coordinate and coordinate squared
operators as:
\begin{equation}
\Delta T_{w}\left( \Phi ;\Psi _{t}\right) =-\frac{i\hbar \Gamma \left[
p_{f}+i\hbar \Gamma x_{f}\right] }{M}\Delta x_{w}\left( \Phi ;\Psi
_{t}\right) -\frac{\hbar ^{2}\Gamma ^{2}}{2M}\Delta x_{w}^{2}\left( \Phi
;\Psi _{t}\right) .  \label{3.14}
\end{equation}%
With some algebra one then finds that
\begin{eqnarray}
\left\langle \Delta T_{w}^{\ast }\left( \Phi ;\Psi _{t}\right) \Delta
x_{w}\left( \Phi ;\Psi _{t}\right) \right\rangle  &=&\frac{i\hbar \Gamma %
\left[ p_{f}-i\hbar \Gamma x_{f}\right] }{M}\left\langle \left\vert \Delta
x_{w}\left( \Phi ;\Psi _{t}\right) \right\vert ^{2}\right\rangle   \notag \\
&-&\frac{\hbar ^{2}\Gamma ^{2}}{2M}\left\langle \left\vert \Delta
x_{w}\left( \Phi ;\Psi _{t}\right) \right\vert ^{2}\Delta x_{w}^{\ast
}\left( \Phi ;\Psi _{t}\right) \right\rangle .  \label{3.15}
\end{eqnarray}%
Using the second equality on the r.h.s of Eq. (\ref{2.19}) the weak value
uncertainty relation takes the form:%
\begin{eqnarray}
&&\frac{M^{2}}{\hbar ^{4}\Gamma ^{4}}\left\langle \left\vert \Delta
T_{w}\left( \Phi ;\Psi _{t}\right) \right\vert ^{2}\right\rangle   \notag \\
&\geq &\left( \frac{p_{f}^{2}}{\hbar ^{2}\Gamma ^{2}}+x_{f}^{2}\right)
\left\langle \left\vert \Delta x_{w}\left( \Phi ;\Psi _{t}\right)
\right\vert ^{2}\right\rangle +\frac{1}{4}\frac{\left\vert \left\langle
\left\vert \Delta x_{w}\left( \Phi ;\Psi _{t}\right) \right\vert ^{2}\Delta
x_{w}^{\ast }\left( \Phi ;\Psi _{t}\right) \right\rangle \right\vert ^{2}}{%
\left\langle \left\vert \Delta x_{w}\left( \Phi ;\Psi _{t}\right)
\right\vert ^{2}\right\rangle }  \notag \\
&&+\frac{p_{f}}{\hbar \Gamma }\left\langle \left\vert \Delta x_{w}\left(
\Phi ;\Psi _{t}\right) \right\vert ^{2}\text{Im}\Delta x_{w}\left( \Phi
;\Psi _{t}\right) \right\rangle -x_{f}\left\langle \left\vert \Delta
x_{w}\left( \Phi ;\Psi _{t}\right) \right\vert ^{2}\text{Re}\Delta
x_{w}\left( \Phi ;\Psi _{t}\right) \right\rangle   \label{3.16}
\end{eqnarray}%
showing that any uncertainty in the time averaged weak value of the position
coordinate leads also to an uncertainty in the kinetic energy. Even if one
chooses the post-selected state to be localized about $x_{f}=p_{f}=0$ one
remains with a nontrivial relation between the two quantities.

\subsection{Weak values and uncertainty in the Stern-Gerlach experiment}

\subsubsection{The model}

The theory of the Stern-Gerlach experiment in which an inhomogeneous field
allows for the strong measurement of the projection of the spin for spin \ $%
1/2$ particles such as electrons or silver atoms has been worked out by
Benitez Rodriguez et al \cite{benitez}. Here, we adapt their results 
to consider weak values of the spin operators in such an experiment. They
considered a particle with mass $M$ which evolves in time under the
Hamiltonian%
\begin{equation}
\hat{H}=\frac{\hat{p}_{r}^{2}}{2M}-\mu _{c}\hat{\sigma}_{z}b\left( y\right) z
\label{3.17}
\end{equation}%
such that the inhomogeneous magnetic field is assumed to be in the $z$
direction only with component%
\begin{equation}
B_{z}=-b\left( y\right) z  \label{3.18}
\end{equation}%
with $b$ denoting the inhomogeneous field strength. The spin states are
eigenvalues of the $\frac{1}{2}\hat{\sigma}_{z}$ spin operator through its Pauli matrix and denoted
as $|\uparrow _{z}\rangle $ and $|\downarrow _{z}\rangle $ with eigenvalues $%
\pm 1/2$ respectively such that in this notation the eigenvalues of $\hat{%
\sigma}_{z}$ are $\pm 1$. The parameter $\mu _{c}=ge\hbar /\left( 2M\right) $ expresses the strength of the interaction. Here, $g$ is the gyromagnetic ratio and $e$ the electron charge. In their
model, the field exists everywhere and therefore is analytically soluble. In
practice, the field is limited to a spatial region such that if the particle
is moving originally in the $y$ direction%
\begin{equation}
b\left( y\right) =b\theta \left( y\right) \theta \left( l-y\right)
\label{3.19}
\end{equation}%
where $\theta \left( x\right) $ is the unit step function. In other words,
the field is limited to the length $l$ in the $y$ direction.

The coupling between the vertical ($z$) and horizontal ($y$) degrees of
freedom makes it in principle impossible to solve the problem exactly.
However, with a "reasonable" approximation, one may still present an
analytic solution to the motion of the particle. Specifically, we will
assume that the pre-selected spatial state of the particle is described by
the coherent state wavepacket
\begin{equation}
\left\langle x,y,z|\psi _{0}\right\rangle =\frac{1}{\left( \pi d^{2}\right)
^{3/4}}\exp \left( -\frac{x^{2}+y^{2}+z^{2}}{2d^{2}}+ik_{y}\left(
y-y_{i}\right) \right)  \label{3.20}
\end{equation}%
where $d$ is the spatial width of the wavepacket and the subscript $0$
denotes the initial time. It is localized about the initial spatial point $%
\left( 0,0,0\right) $ with mean momentum in the $y$ direction $\hbar k_{y}$. The width parameter $%
d$ will be chosen large enough such that the momentum spread $\left(
dk_{y}\right) ^{-1}\ll 1$. With these conditions, we may consider a
classical path approximation by which the evolution of the motion in the $y$
direction may be considered to be classical. This implies that the time
during which the particle feels the magnetic field is well approximated as
\begin{equation}
\tau =\frac{Ml}{\hbar k_{y}}.  \label{3.21}
\end{equation}

The normalized pre-selected state is then described by the general form
(with the superposition coefficients $\left\vert \alpha _{i}\right\vert
^{2}+\left\vert \beta _{i}\right\vert ^{2}=1$) as%
\begin{equation}
|\Psi _{0}\rangle =|\psi _{0}\rangle \left[ \alpha _{i}|\uparrow _{z}\rangle
+\beta _{i}|\downarrow _{z}\rangle \right] .  \label{3.22}
\end{equation}%
The time dependent width is given by%
\begin{equation}
d_{t}^{2}=d^{2}+\frac{i\hbar t}{M}.  \label{3.23}
\end{equation}%
We then use reduced variables such that all lengths ($x,y,z,l$) are reduced according to
\begin{equation}
x/d\rightarrow x,y/d\rightarrow y,z/d\rightarrow z,l/d\rightarrow l.
\label{3.24}
\end{equation}%
and the  time and wavenumbers as%
\begin{equation}
\frac{\hbar t}{Md^{2}}\rightarrow t,k_{y}d\rightarrow k_{y}.  \label{3.25}
\end{equation}%
The physical field strength parameter $\alpha =\tau \mu _{c}b$ has dimension
of $\hbar /d$ so it reduces naturally to%
\begin{equation}
\frac{\alpha d}{\hbar }\rightarrow \alpha .  \label{3.26}
\end{equation}%

With these preliminaries one finds that the time dependent wavefunction at a
position $y\gg l/2$, that is, to the right of the region where the particle
feels the magnetic field is
\begin{equation}
\left\langle {x,y,z}\mathbf{|}\Psi _{t}\right\rangle =\alpha _{i}\phi
_{+}\left( x,y,z;t\right) |\uparrow _{z}\rangle +\beta _{i}\phi _{-}\left(
x,y,z;t\right) |\downarrow _{z}\rangle  \label{3.27}
\end{equation}%
with%
\begin{eqnarray}
&&\phi _{+}\left( x,y\gg l/2,z;t\right) =\left[ \frac{1}{\left( 1+it\right)
\sqrt{\pi }}\right] ^{3/2}  \notag \\
&&\cdot \exp \left( -i\frac{\alpha ^{2}t}{6}-\frac{x^{2}+\left(
y-ik_{y}\right) ^{2}+\left( z-\frac{1}{2}\alpha t\right) ^{2}}{2\left(
1+it\right) }-\frac{k_{y}^{2}}{2}-i\alpha z\right)  \label{3.28}
\end{eqnarray}
and
\begin{eqnarray}
\phi _{-}\left( x,y\gg l/2,z;t\right) =\phi _{+}\left( x,y\gg
l/2,z;t\right) \exp \left( 2i\alpha z-\frac{\alpha zt}{\left( 1+it\right) }%
\right).  \label{3.29}
\end{eqnarray}%
The two spatial functions are normalized, that is $%
\left\langle \phi _{+}|\phi _{+}\right\rangle =\left\langle \phi _{-}|\phi
_{-}\right\rangle =1$ but are not orthogonal:%
\begin{equation}
\left\langle \phi _{+}|\phi _{-}\right\rangle =\exp \left( -\alpha
^{2}\left( 1+\frac{9}{4}t^{2}\right) \right) .  \label{3.30}
\end{equation}%
If the field is very strong or at long times the overlap will
effectively vanish, implying a full spatial separation of the up and down
spin beams.

\subsubsection{Strong value (standard)\ uncertainty relation}

Before dealing with weak values, we consider the time dependent strong values
for the three spins. By definition, the strong value of the $j$-th
component of the Pauli matrix at time $t$ is:%
\begin{equation}
S_{j,s}\left( t\right) =\left\langle \Psi _{t}\left\vert \hat{\sigma}%
_{j}\right\vert \Psi _{t}\right\rangle ,\text{ \ \ }j=x,y,z.  \label{3.31}
\end{equation}
Due to the fact that the two spin states are orthogonal to each other we
have that $\left\langle \Psi _{t}|\Psi _{t}\right\rangle =1$ and the time
dependent strong values for the three spins are:
\begin{eqnarray}
S_{x,s}\left( t\right) &=&\left\langle \phi _{-}|\phi _{+}\right\rangle
\left( \beta _{i}^{\ast }\alpha _{i}+\alpha _{i}^{\ast }\beta _{i}\right)
\label{3.32} \\
S_{y,s}\left( t\right) &=&i\left\langle \phi _{-}|\phi _{+}\right\rangle
\left( \beta _{i}^{\ast }\alpha _{i}-\alpha _{i}^{\ast }\beta _{i}\right)
\label{3.33} \\
S_{z,s}\left( t\right) &=&\left\vert \alpha _{i}\right\vert ^{2}-\left\vert
\beta _{i}\right\vert ^{2}.  \label{3.34}
\end{eqnarray}%
If, for example, we choose $\alpha _{i}=\frac{1}{\sqrt{2}},\beta _{i}=\frac{1%
}{\sqrt{2}}\exp \left( i\chi _{i}\right) $, then the mean value of the
vertical spin vanishes as expected. Since we have not post-selected any
state, the positive and negative spin beams have the same probability and
the mean value of the vertical spin vanishes. If we choose the phase $\chi
_{i}=0,\pi $, then the initial state is an eigenstate of the $\hat{\sigma}%
_{x}$ operator and the strong value of the spin in the $x$ direction is
equal to $\left\langle \phi _{-}|\phi _{+}\right\rangle $.

The (cyclic) commutation and anti-commutation relations of the Pauli matrices
are:%
\begin{equation}
\left[ \hat{\sigma}_{x,}\hat{\sigma}_{y}\right] =2i\hat{\sigma}_{z},\text{ \
\ }\left\{ \hat{\sigma}_{x,}\hat{\sigma}_{y}\right\} =0.  \label{3.35}
\end{equation}%
The strong value (Robertson) uncertainty relation is then
\begin{eqnarray}
\left\langle \Psi _{t}\left\vert \Delta \hat{\sigma}_{x}^{2}\right\vert \Psi
_{t}\right\rangle \left\langle \Psi _{t}\left\vert \Delta \hat{\sigma}%
_{y}^{2}\right\vert \Psi _{t}\right\rangle &=&\left[ \left\langle \Psi
_{t}\left\vert \hat{\sigma}_{x}^{2}\right\vert \Psi _{t}\right\rangle
-\left\langle \Psi _{t}\left\vert \hat{\sigma}_{x}\right\vert \Psi
_{t}\right\rangle ^{2}\right] \left[ \left\langle \Psi _{t}\left\vert \hat{%
\sigma}_{y}^{2}\right\vert \Psi _{t}\right\rangle -\left\langle \Psi
_{t}\left\vert \hat{\sigma}_{y}\right\vert \Psi _{t}\right\rangle ^{2}\right]
\notag \\
&\geq &-\frac{1}{4}\left\langle \Psi _{t}\left\vert \left[ \hat{\sigma}_{x},%
\hat{\sigma}_{y}\right] \right\vert \Psi _{t}\right\rangle ^{2}=\left\langle
\Psi _{t}\left\vert \hat{\sigma}_{z}\right\vert \Psi _{t}\right\rangle ^{2}.
\label{3.36}
\end{eqnarray}%
Using the notation
\begin{equation}
\beta _{i}=\beta \exp \left( ib\right) ,\alpha _{i}=\alpha \exp \left(
ia\right)  \label{3.37}
\end{equation}%
one has that:%
\begin{eqnarray}
\left\langle \Psi _{t}\left\vert \Delta \hat{\sigma}_{x}^{2}\right\vert \Psi
_{t}\right\rangle &=&1-4\alpha ^{2}\beta ^{2}\left\langle \phi _{-}|\phi
_{+}\right\rangle ^{2}\cos ^{2}\left( a-b\right)  \label{3.38} \\
\left\langle \Psi _{t}\left\vert \Delta \hat{\sigma}_{y}^{2}\right\vert \Psi
_{t}\right\rangle &=&1-4\alpha ^{2}\beta ^{2}\left\langle \phi _{-}|\phi
_{+}\right\rangle ^{2}\sin ^{2}\left( a-b\right)  \label{3.39} \\
\left\langle \Psi _{t}\left\vert \hat{\sigma}_{z}\right\vert \Psi
_{t}\right\rangle ^{2} &=&\left( \alpha ^{2}-\beta ^{2}\right) ^{2}.
\label{3.40}
\end{eqnarray}%
The uncertainty relation Eq. (\ref{3.36}) then takes the form:%
\begin{eqnarray}
&&\left\langle \Psi _{t}\left\vert \Delta \hat{\sigma}_{x}^{2}\right\vert
\Psi _{t}\right\rangle \left\langle \Psi _{t}\left\vert \Delta \hat{\sigma}%
_{y}^{2}\right\vert \Psi _{t}\right\rangle -\left\langle \Psi _{t}\left\vert
\hat{\sigma}_{z}\right\vert \Psi _{t}\right\rangle ^{2}  \notag \\
&=&4\alpha ^{2}\beta ^{2}\left( 1-\left\langle \phi _{-}|\phi
_{+}\right\rangle ^{2}+\left\langle \phi _{-}|\phi _{+}\right\rangle
^{4}\sin ^{2}\left[ 2\left( a-b\right) \right] \right) \geq 0.  \label{3.41}
\end{eqnarray}%
and one notes that the r.h.s. is positive, since $\left\langle \phi
_{-}|\phi _{+}\right\rangle \leq 1$.

\subsubsection{Weak value uncertainty relation}

We now want to study the time averaged weak value uncertainty relation. For
this purpose, one must choose a post-selected state. We assume a "screen" at $%
y=y_{s}\gg l/2$, that is, the post-selected coordinate in the $y$ direction
is chosen far to the right of the applied field. The spatial post-selected
state in the $x$ and $z$ direction will be denoted as $|\phi _{f}\left(
x,z\right) \rangle $ so that the post-selected state takes the form%
\begin{equation}
|\Phi _{f}\rangle =|\phi _{f},y_{s}\rangle \left[ \alpha _{f}|\uparrow
_{z}\rangle +\beta _{f}|\downarrow _{z}\rangle \right] .  \label{3.42}
\end{equation}
Reminding ourselves that the three spin operators $\hat{\sigma}_{x},\hat{%
\sigma}_{y}$ and $\hat{\sigma}_{z}$ are defined such that
\begin{eqnarray}
\hat{\sigma}_{x}| &\uparrow &_{z}\rangle =|\downarrow _{z}\rangle ,\hat{%
\sigma}_{x}|\downarrow _{z}\rangle =|\uparrow _{z}\rangle  \label{3.43} \\
\hat{\sigma}_{y}| &\uparrow &_{z}\rangle =i|\downarrow _{z}\rangle ,\hat{%
\sigma}_{y}|\downarrow _{z}\rangle =-i|\uparrow _{z}\rangle  \label{3.44} \\
\hat{\sigma}_{z}| &\uparrow &_{z}\rangle =|\uparrow _{z}\rangle ,\hat{\sigma}%
_{z}|\downarrow _{z}\rangle =-|\downarrow _{z}\rangle ,  \label{3.45}
\end{eqnarray}%
and using the notation%
\begin{equation}
R\left( y_{s}\right) =\frac{\langle \phi _{f},y_{s}|\phi _{-}\rangle }{%
\langle \phi _{f},y_{s}|\phi _{+}\rangle }  \label{3.46}
\end{equation}%
for the ratio of the two overlaps, we find the following formal results
for the post-selected weak values of the three spin operators%
\begin{eqnarray}
S_{x,w} &=&\frac{\left\langle \Phi _{f}\left\vert \hat{\sigma}%
_{x}\right\vert \Psi _{t}\right\rangle }{\left\langle \Phi _{f}|\Psi
_{t}\right\rangle }=\frac{\beta _{f}^{\ast }\alpha _{i}+\alpha _{f}^{\ast
}\beta _{i}R\left( y_{s}\right) }{\alpha _{f}^{\ast }\alpha _{i}+\beta
_{f}^{\ast }\beta _{i}R\left( y_{s}\right) }  \label{3.47} \\
S_{y,w} &=&\frac{\left\langle \Phi _{f}\left\vert \hat{\sigma}%
_{y}\right\vert \Psi _{t}\right\rangle }{\left\langle \Phi _{f}|\Psi
_{t}\right\rangle }=i\frac{\beta _{f}^{\ast }\alpha _{i}-\alpha _{f}^{\ast
}\beta _{i}R\left( y_{s}\right) }{\alpha _{f}^{\ast }\alpha _{i}+\beta
_{f}^{\ast }\beta _{i}R\left( y_{s}\right) }  \label{3.48} \\
S_{z,w} &=&\frac{\left\langle \Phi _{f}\left\vert \hat{\sigma}%
_{z}\right\vert \Psi _{t}\right\rangle }{\left\langle \Phi _{f}|\Psi
_{t}\right\rangle }=\frac{\alpha _{f}^{\ast }\alpha _{i}-\beta _{f}^{\ast
}\beta _{i}R\left( y_{s}\right) }{\alpha _{f}^{\ast }\alpha _{i}+\beta
_{f}^{\ast }\beta _{i}R\left( y_{s}\right) }.  \label{3.49}
\end{eqnarray}%

One may then re-express the overlap ratio $R\left( y_{s}\right) $ in terms
of the weak spin value in the $z$ direction
\begin{equation}
R\left( y_{s}\right) =\frac{\alpha _{f}^{\ast }\alpha _{i}\left[ 1-S_{z,w}%
\right] }{\beta _{f}^{\ast }\beta _{i}\left[ 1+S_{z,w}\right] }  \label{3.50}
\end{equation}%
to find the linear relations between the weak values in the $x$ and $y$
directions and the vertical direction:%
\begin{equation}
S_{x,w}=\frac{1}{2}\left[ \left( \frac{\beta _{f}^{\ast }}{\alpha _{f}^{\ast
}}+\frac{\alpha _{f}^{\ast }}{\beta _{f}^{\ast }}\right) +\left( \frac{\beta
_{f}^{\ast }}{\alpha _{f}^{\ast }}-\frac{\alpha _{f}^{\ast }}{\beta
_{f}^{\ast }}\right) S_{z,w}\right]  \label{3.51}
\end{equation}
\begin{equation}
S_{y,w}=-\frac{i}{2}\left[ \left( \frac{\alpha _{f}^{\ast }}{\beta
_{f}^{\ast }}-\frac{\beta _{f}^{\ast }}{\alpha _{f}^{\ast }}\right) -\left(
\frac{\alpha _{f}^{\ast }}{\beta _{f}^{\ast }}+\frac{\beta _{f}^{\ast }}{%
\alpha _{f}^{\ast }}\right) S_{z,w}\right] .  \label{3.52}
\end{equation}%
This allows us to derive a linear relation for the fluctuations of the
spins:
\begin{equation}
\Delta S_{x,w}=\frac{1}{2}\left( \frac{\beta _{f}^{\ast }}{\alpha _{f}^{\ast
}}-\frac{\alpha _{f}^{\ast }}{\beta _{f}^{\ast }}\right) \Delta S_{z,w}
\label{3.53}
\end{equation}
\begin{equation}
\Delta S_{y,w}=\frac{i}{2}\left( \frac{\alpha _{f}^{\ast }}{\beta _{f}^{\ast
}}+\frac{\beta _{f}^{\ast }}{\alpha _{f}^{\ast }}\right) \Delta S_{z,w},
\label{3.54}
\end{equation}%
and therefore
\begin{equation}
\Delta S_{x,w}=i\frac{\left( \alpha _{f}^{\ast 2}-\beta _{f}^{\ast 2}\right)
}{\left( \alpha _{f}^{\ast 2}+\beta _{f}^{\ast 2}\right) }\Delta S_{y,w}.
\label{3.55}
\end{equation}%
As in the case of the position and momentum operators, this linear relation
implies that the weak value uncertainty principle does not pose any
restriction on the weak values of the operators $\sigma _{x}$ and $\sigma
_{y}$ even though they do not commute.

The weak value time averaged commutator and anti-commutator are found to be:%
\begin{equation}
\left\langle \left[ \Delta S_{x,w}^{\dag },\Delta S_{y,w}\right]
\right\rangle =\frac{i}{2}\left\langle \left\vert \Delta S_{z,w}\right\vert
^{2}\right\rangle \left[ \frac{\left\vert \beta _{f}\right\vert
^{2}-\left\vert \alpha _{f}\right\vert ^{2}}{\left\vert \alpha
_{f}\right\vert ^{2}\left\vert \beta _{f}\right\vert ^{2}}\right]
\label{3.56}
\end{equation}%
\begin{equation}
\left\langle \left\{ \Delta S_{x,w}^{\dag },\Delta S_{y,w}\right\}
\right\rangle =\frac{i}{2}\left\langle \left\vert \Delta S_{z,w}\right\vert
^{2}\right\rangle \left[ \frac{\beta _{f}^{2}\alpha _{f}^{\ast 2}-\alpha
_{f}^{2}\beta _{f}^{\ast 2}}{\left\vert \alpha _{f}\right\vert
^{2}\left\vert \beta _{f}\right\vert ^{2}}\right] .  \label{3.57}
\end{equation}%
These are quite different from their strong value analogues as derived from
Eqs. (\ref{3.35}). The weak value commutator may vanish, if the post selected
state is chosen (as is often the case) such that $\left\vert \beta
_{f}\right\vert ^{2}=\left\vert \alpha _{f}\right\vert ^{2}$. Similarly, the
weak value anti-commutator need not vanish. With these results it is also
straightforward to find that
\begin{equation}
\left\langle \left\vert \Delta S_{x,w}\right\vert ^{2}\right\rangle
\left\langle \left\vert \Delta S_{y,w}\right\vert ^{2}\right\rangle =\frac{%
\left\vert \left( \beta _{f}^{4}-\alpha _{f}^{4}\right) \right\vert ^{2}}{%
16\left\vert \alpha _{f}\beta _{f}\right\vert ^{4}}\left\langle \left\vert
\Delta S_{z,w}\right\vert ^{2}\right\rangle ^{2},  \label{3.58}
\end{equation}%
showing that if the standard deviation from the time averaged weak value of
the vertical spin vanishes, then there is no constraint on the analogous
standard deviation of the weak values in the $x$ and \ $y$ directions.

\subsubsection{The time-spin uncertainty relation}

It is also illuminating to study the relationship between the time
averaged variance of the weak value of any of the spins and the variance of
the time it takes the particle passing through the Stern-Gerlach machine to
reach the final screen. Here, the weak value of the time is defined as in
Eq. (\ref{3.1}) and the weak values of the spin as in Eqs. (\ref{3.47})-(\ref
{3.49}). To apply the weak value uncertainty principle as in Eq. (\ref{2.19})
we first note that the time probability distribution at the post-selected
state as chosen in Eq. (\ref{3.42}) is%
\begin{eqnarray}
&&P_{\Psi }\left( t;\Phi \right) =\notag \\ &&\frac{\left\vert \alpha _{f}^{\ast }\alpha
_{i}\int_{-\infty }^{\infty }dxdz\phi _{f}^{\ast }\left( x,z,y_{s}\right)
\phi _{+}\left( x,y_{s},z;t\right) +\beta _{f}^{\ast }\beta
_{i}\int_{-\infty }^{\infty }dxdz\phi _{f}^{\ast }\left( x,z,y_{s}\right)
\phi _{-}\left( x,y_{s},z;t\right) \right\vert ^{2}}{N\left( \Phi ;\Psi
\right) }. \notag \\ \label{3.59}
\end{eqnarray}%
The time-spin uncertainty relation is then readily seen to be:
\begin{equation}
\left\langle \left\vert \Delta S_{j,w}\right\vert ^{2}\right\rangle
\left\langle \left\vert \Delta t\right\vert ^{2}\right\rangle \geq
\left\vert \left\langle tS_{j,w}\right\rangle \right\vert ^{2}, \label{3.60}
\end{equation}%
with $j=x,y,z$. This is a non-trivial statement since it shows that the variance of the weak
spin is directly proportional to its time modulated magnitude. In other
words, large anomalous values of the spin lead to large variance of the weak
spin values and thus imply larger experimental averaging needed than for
"normal" spin values.

\renewcommand{\theequation}{4.\arabic{equation}} \setcounter{section}{3} %
\setcounter{equation}{0}

\section{Specific Applications}

\subsection{Potential scattering}

\subsubsection{The coordinate-momentum uncertainty}

To exemplify the consequences of the weak value uncertainty principle, we
consider in some detail the scattering of a particle through a potential
barrier, where for the sake of simplicity, the potential $V\left( q\right) $
vanishes as $q\rightarrow \pm \infty $. For an incident momentum $p_{i}$ of
a particle with mass $M$ from the left, the transmission amplitude through
the barrier is denoted by $T\left( p_{i}\right) $. We choose the pre-
selected state to be a coherent state localized about a point $x_{i}$
sufficiently far to the left of the potential such that its overlap with the
potential is negligible:%
\begin{equation}
\left\langle x|\Psi \right\rangle =\left( \frac{\Gamma }{\pi }\right)
^{1/4}\exp \left( -\frac{\Gamma }{2}\left( x-x_{i}\right) ^{2}+\frac{i}{%
\hbar }p_{i}\left( x-x_{i}\right) \right) .  \label{4.1}
\end{equation}%
This implies that $\Gamma x_{i}^{2}\gg 1$. As shown in \cite{eli2} the time evolved
wavefunction in the region far to the left of the potential may be
approximated by steepest descent to be:%
\begin{eqnarray}
\left\langle x|\Psi \left( t\right) \right\rangle &\simeq &\left( \frac{%
\Gamma M^{2}}{\pi \left( M+it\hbar \Gamma \right) ^{2}}\right) ^{\frac{1}{4}%
}T\left( \frac{Mp_{i}-i\hbar \Gamma M\left( x_{i}-x\right) }{\left(
M+it\hbar \Gamma \right) }\right)  \notag \\
&\cdot &\exp \left( -\frac{p_{i}^{2}}{2\hbar ^{2}\Gamma }+\frac{M\Gamma }{2}%
\frac{\left( i\left( x_{i}-x\right) -\frac{p_{i}}{\hbar \Gamma }\right) ^{2}%
}{\left[ M+it\hbar \Gamma \right] }\right) .  \label{4.2}
\end{eqnarray}%
We then choose the post-selected state to be a Gaussian localized about the
point $x_{f}$ sufficiently far out to the left of the potential so that it
does not overlap with the potential, with vanishing mean momentum and to
simplify the width parameter $\Gamma _{f}$ $\gg \Gamma $ is chosen
sufficiently large such that the post-selected state effectively localizes
the particle at the point $x_{f}$:
\begin{equation}
\left\langle x|\Phi _{f}\right\rangle =\left( \frac{\Gamma _{f}}{\pi }%
\right) ^{1/4}\exp \left( -\frac{\Gamma _{f}}{2}\left( x-x_{f}\right)
^{2}\right) .  \label{4.3}
\end{equation}%
Within the steepest descent approximation, using the notation
\begin{equation}
\Delta x=x_{f}-x_{i}  \label{4.4}
\end{equation}%
the weak value of the momentum is%
\begin{equation}
p_{w}\left( \Phi _{f};\Psi _{t}\right) \simeq \frac{M^{2}p_{i}+\hbar
^{2}\Gamma ^{2}M\Delta xt}{\left[ M^{2}+t^{2}\hbar ^{2}\Gamma ^{2}\right] }%
+i\hbar M\Gamma \frac{p_{i}t-M\Delta x}{\left[ M^{2}+t^{2}\hbar ^{2}\Gamma
^{2}\right] }+O\left( \frac{\Gamma }{\Gamma _{f}}\right)  \label{4.5}
\end{equation}
and the weak value of the coordinate (see Eq. \ref{3.10}) is:
\begin{equation}
x_{w}\left( \Phi ;\Psi _{t}\right) =x_{f}+O\left( \frac{\Gamma }{\Gamma _{f}}%
\right) .  \label{4.6}
\end{equation}%
Since the weak value of the coordinate is time independent, it has no
uncertainty associated with it, therefore the variance of the weak value of
the momentum can, by suitable choice of the width parameter of the pre-
selected state, be made arbitrarily small.

The steepest descent approximation for the time probability distribution is
readily found to be (ignoring terms of the order of $\frac{\Gamma }{\Gamma
_{f}}$):
\begin{equation}
\frac{\left\vert \left\langle \Phi _{f}|\Psi \left( t\right) \right\rangle
\right\vert ^{2}}{\int_{0}^{\infty }dt\left\vert \left\langle \Phi _{f}|\Psi
\left( t\right) \right\rangle \right\vert ^{2}}\simeq p_{i}\left( \frac{%
p_{i}^{2}\Gamma }{\pi \left[ M^{2}p_{i}^{2}+M^{2}\Delta x^{2}\hbar
^{2}\Gamma ^{2}\right] }\right) ^{\frac{1}{2}}\exp \left( -\frac{%
p_{i}^{2}\Gamma \left( M\Delta x-p_{i}t\right) ^{2}}{\left[
M^{2}p_{i}^{2}+M^{2}\Delta x^{2}\hbar ^{2}\Gamma ^{2}\right] }\right)
\label{4.7}
\end{equation}%
so that the time averaged mean of the momentum is within the steepest
descent approximation:%
\begin{equation}
\left\langle p_{w}\left( \Phi _{f};\Psi _{t}\right) \right\rangle \simeq
p_{i}+O\left( \frac{\Gamma }{\Gamma _{f}}\right)  \label{4.8}
\end{equation}%
while the time averaged mean of the weak value of the coordinate is of
course:%
\begin{equation}
\left\langle x_{w}\left( \Phi ;\Psi _{t}\right) \right\rangle \simeq
x_{f}+O\left( \frac{\Gamma }{\Gamma _{f}}\right) .  \label{4.9}
\end{equation}%
The variance of the weak value of the momentum is readily seen to be:%
\begin{equation}
\left\langle \left\vert p_{w}\left( \Phi _{f};\Psi _{t}\right) \right\vert
^{2}\right\rangle -\left\vert \left\langle p_{w}\left( \Phi _{f};\Psi
_{t}\right) \right\rangle \right\vert ^{2}\simeq \frac{\hbar ^{2}\Gamma }{2}
\label{4.10}
\end{equation}%
and as noted before can be made arbitrarily small by choosing the width parameter of
the initial state to be sufficiently small.

\subsubsection{The kinetic energy uncertainty}

Within the same framework as above, the weak value of the kinetic energy is
readily seen to be:%
\begin{equation}
T_{w}\left( \Phi _{f};\Psi _{t}\right) \simeq \frac{1}{2}\left[ \frac{\hbar
^{2}\Gamma }{\left[ M+it\hbar \Gamma \right] }-\frac{M\left( ip_{i}-\hbar
\Gamma \Delta x\right) ^{2}}{\left[ M+it\hbar \Gamma \right] ^{2}}\right]
\label{4.11}
\end{equation}%
and its time averaged mean is thus complex:%
\begin{equation}
\left\langle T_{w}\left( \Phi _{f};\Psi _{t}\right) \right\rangle \simeq
\frac{p_{i}^{2}}{2M}\left[ 1+\frac{\hbar ^{2}\Gamma }{\left[
p_{i}^{2}+\Delta x^{2}\hbar ^{2}\Gamma ^{2}\right] }\right] -\frac{i\hbar
p_{i}\hbar ^{2}\Gamma ^{2}\Delta x}{M\left[ p_{i}^{2}+\Delta x^{2}\hbar
^{2}\Gamma ^{2}\right] }.  \label{4.12}
\end{equation}%
In the limit that the distance between the initial and final points becomes
very large, such that $\Gamma \Delta x^{2}\gg p_{i}^{2}/\left( \hbar
^{2}\Gamma \right) $ and $\Gamma \Delta x^{2}\gg 1$ one readily finds that%
\begin{equation}
\left\langle T_{w}\left( \Phi _{f};\Psi _{t}\right) \right\rangle \simeq
\frac{p_{i}^{2}}{2M}\left[ 1+\frac{1}{\Delta x^{2}\Gamma }\right] -\frac{i%
\sqrt{\hbar ^{2}\Gamma }p_{i}}{M\sqrt{\Gamma \Delta x^{2}}}+O\left( \frac{%
p_{i}^{2}}{\sqrt{\Gamma \Delta x^{2}}\Delta x^{2}\hbar ^{2}\Gamma ^{2}}%
\right) .  \label{4.13}
\end{equation}%
With some further algebra one also finds that within the steepest descent
estimate the variance of the weak value of the kinetic energy is%
\begin{eqnarray}
&&\left\langle \left\vert T_{w}\left( \Phi _{f};\Psi _{t}\right) \right\vert
^{2}\right\rangle -\left\vert \left\langle T_{w}\left( \Phi _{f};\Psi
_{t}\right) \right\rangle \right\vert ^{2}  \notag \\
&\simeq &\frac{2\hbar ^{2}\Gamma p_{i}^{6}\left( \Delta x^{2}\hbar
^{2}\Gamma ^{2}\right) }{M^{2}\left[ p_{i}^{2}+\Delta x^{2}\hbar ^{2}\Gamma
^{2}\right] ^{3}}+\frac{\hbar ^{2}\Gamma \left( \hbar ^{2}\Gamma
+2p_{i}^{2}\right) ^{2}}{8M^{2}\left[ p_{i}^{2}+\Delta x^{2}\hbar ^{2}\Gamma
^{2}\right] }+\frac{\hbar ^{6}\Gamma ^{3}\Gamma \Delta x^{2}p_{i}^{2}}{2M^{2}%
\left[ p_{i}^{2}+\Delta x^{2}\hbar ^{2}\Gamma ^{2}\right] ^{2}}.
\label{4.14}
\end{eqnarray}%
As before in the limit that $\Gamma \Delta x^{2}\gg p_{i}^{2}/\left( \hbar
^{2}\Gamma \right) $ and $\Gamma \Delta x^{2}\gg 1$ this simplifies to
\begin{equation}
\left\langle \left\vert T_{w}\left( \Phi _{f};\Psi _{t}\right) \right\vert
^{2}\right\rangle -\left\vert \left\langle T_{w}\left( \Phi _{f};\Psi
_{t}\right) \right\rangle \right\vert ^{2}\simeq \frac{1}{\Gamma \Delta x^{2}%
}\left( \frac{\hbar ^{4}\Gamma ^{2}}{8M^{2}}+\frac{p_{i}^{4}}{2M^{2}}+\frac{%
\hbar ^{2}\Gamma p_{i}^{2}}{M^{2}}\right)  \label{4.15}
\end{equation}%
demonstrating that one should expect that the variance in the kinetic energy
vanishes when the distance between the initial and final points is
sufficiently large.

It remains to consider the mean time and its variance. Within the steepest
descent approximation one has that the mean time and its variance are given
by:
\begin{eqnarray}
\left\langle t\right\rangle &=&\frac{M\Delta x}{p_{i}}  \label{4.16} \\
\left\langle \left( t-\left\langle t\right\rangle \right) ^{2}\right\rangle
&=&\frac{\left[ M^{2}p_{i}^{2}+M^{2}\Delta x^{2}\hbar ^{2}\Gamma ^{2}\right]
}{2\Gamma p_{i}^{4}}  \label{4.17}
\end{eqnarray}%
and one notes that when $\Gamma \Delta x^{2}\gg p_{i}^{2}/\left(
\hbar ^{2}\Gamma \right) $ and $\Gamma \Delta x^{2}\gg 1$, the time variance
diverges. The product of the kinetic energy and time variances is:%
\begin{eqnarray}
&&\left[ \left\langle \left\vert T_{w}\left( \Phi _{f};\Psi _{t}\right)
\right\vert ^{2}\right\rangle -\left\vert \left\langle T_{w}\left( \Phi
_{f};\Psi _{t}\right) \right\rangle \right\vert ^{2}\right] \left\langle
\left( t-\left\langle t\right\rangle \right) ^{2}\right\rangle  \notag \\
&\simeq &\frac{\hbar ^{2}p_{i}^{2}\left( \Delta x^{2}\hbar ^{2}\Gamma
^{2}\right) }{\left[ p_{i}^{2}+\Delta x^{2}\hbar ^{2}\Gamma ^{2}\right] ^{2}}%
+\frac{\hbar ^{2}\left( \hbar ^{2}\Gamma +2p_{i}^{2}\right) ^{2}}{16p_{i}^{4}%
}+\frac{\hbar ^{6}\Gamma ^{3}\Delta x^{2}}{4p_{i}^{2}\left[ p_{i}^{2}+\Delta
x^{2}\hbar ^{2}\Gamma ^{2}\right] }.  \label{4.18}
\end{eqnarray}%
and in the limit that $\Gamma \Delta x^{2}\gg p_{i}^{2}/\left( \hbar
^{2}\Gamma \right) $ and $\Gamma \Delta x^{2}\gg 1$ we find that%
\begin{eqnarray}
&&\left[ \left\langle \left\vert T_{w}\left( \Phi _{f};\Psi _{t}\right)
\right\vert ^{2}\right\rangle -\left\vert \left\langle T_{w}\left( \Phi
_{f};\Psi _{t}\right) \right\rangle \right\vert ^{2}\right] \left\langle
\left( t-\left\langle t\right\rangle \right) ^{2}\right\rangle  \label{4.19}
\\
&\simeq &\frac{\hbar ^{2}}{4}\left( 1+\frac{2\hbar ^{2}\Gamma }{p_{i}^{2}}+%
\frac{\hbar ^{4}\Gamma ^{2}}{4p_{i}^{4}}\right) \geq \frac{\hbar ^{2}}{4}
\notag
\end{eqnarray}%
in accordance with the weak value time-energy uncertainty relation.

In summary, from the analysis of time averaged weak values one concludes
that in a scattering experiment, one may know the precise value of the
momentum and the kinetic energy of a particle as it hits the screen, that is,
at a known location. However, as the precision increases, the variance in
the time of arrival of the particle at the screen increases, and the product
of the kinetic energy and time variances is limited by the "standard"
uncertainty relation.

\subsection{The Stern-Gerlach experiment}

To demonstrate the result more clearly for the Stern-Gerlach setup we
present some numerical results. For this purpose we have to specify the
post-selected state. The field splits the beam into two parts, one localized
on the positive $z$ axis the other on the negative. We will choose the
spatial part of the post-selected state to be localized on the positive $z$
axis at $z=z_{0}$ where $z_{0}$ is at the center of the positive beam found
on a screen located at and about $x=0$. \ We then choose the final spatial
state to be a Gaussian centered about $z_{0}$:
\begin{equation}
\left\langle x,z|\phi _{f}\right\rangle =\left( \frac{1}{\pi }\right)
^{1/4}\exp \left( -\frac{x^{2}+\left( z-z_{0}\right) ^{2}}{2}\right) .
\label{4.20}
\end{equation}%
The overlap integrals involve a Gaussian integration over the $x$ and $z$
coordinates. One readily finds:%
\begin{eqnarray}
\langle \phi _{f},y_{s}|\phi _{+}\rangle &=&\frac{2}{\sqrt{1+it}\left(
2+it\right) }  \notag \\
&&\cdot \exp \left( -i\frac{\alpha ^{2}t}{6}-\frac{\left( y-ik_{y}\right)
^{2}}{2\left( 1+it\right) }-\frac{\left( z_{0}-\frac{1}{2}\alpha t-i\alpha
\right) ^{2}}{2\left( 2+it\right) }-\frac{k_{y}^{2}+\alpha ^{2}}{2}-i\alpha
z_{0}\right)  \notag \\
&&  \label{4.21}
\end{eqnarray}%
and
\begin{equation}
R\left( y_{s}\right) =\exp \left( i\eta _{t}-Y_{t}\right) ,\eta _{t}=\frac{%
\alpha z_{0}\left( 1+\frac{3}{4}t^{2}\right) }{\left( 1+\frac{t^{2}}{4}%
\right) },Y_{t}=\frac{\alpha z_{0}t}{\left( 1+\frac{t^{2}}{4}\right) }.
\label{4.22}
\end{equation}%
The time dependent densities are then:%
\begin{eqnarray}
\left\vert \langle \phi _{f},y_{s}|\phi +\rangle \right\vert ^{2} &=&\frac{1%
}{\sqrt{1+t^{2}}\left( 1+\frac{t^{2}}{4}\right) }\exp \left( -\frac{\left(
y_{s}-k_{y}t\right) ^{2}}{\left( 1+t^{2}\right) }-\frac{\left( z_{0}-\alpha
t\right) ^{2}}{2\left( 1+\frac{t^{2}}{4}\right) }-\frac{\alpha ^{2}}{2}%
\right)  \label{4.23} \\
\left\vert \langle \phi _{f},y_{s}|\phi -\rangle \right\vert ^{2}
&=&\left\vert \langle \phi _{f},y_{s}|\phi +\rangle \right\vert ^{2}\exp
\left( -\frac{2\alpha tz_{0}}{\left( 1+\frac{t^{2}}{4}\right) }\right).
\label{4.24}
\end{eqnarray}%
\begin{figure}[tbp]
\begin{centering}
\includegraphics[scale=0.4]{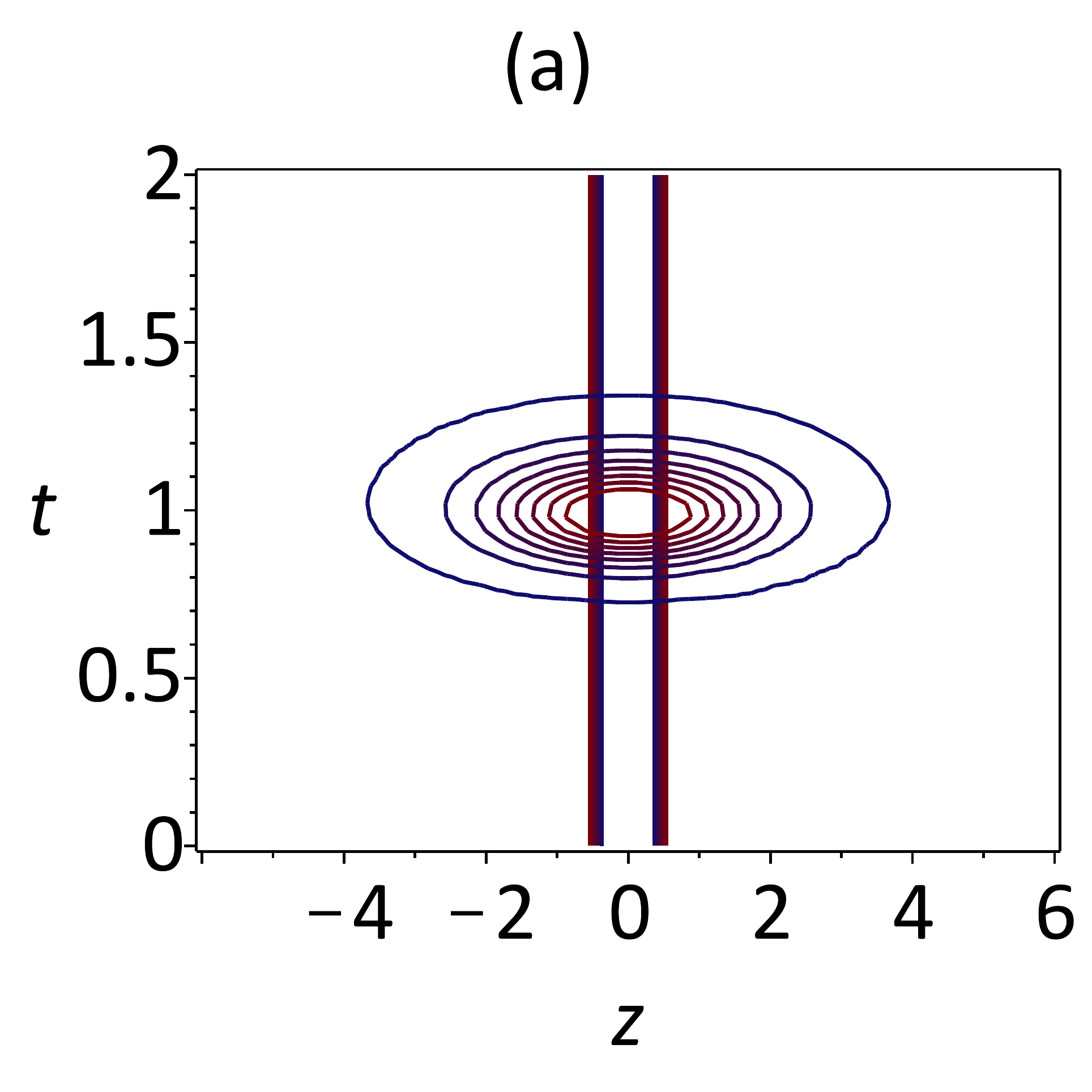}
\includegraphics[scale=0.4]{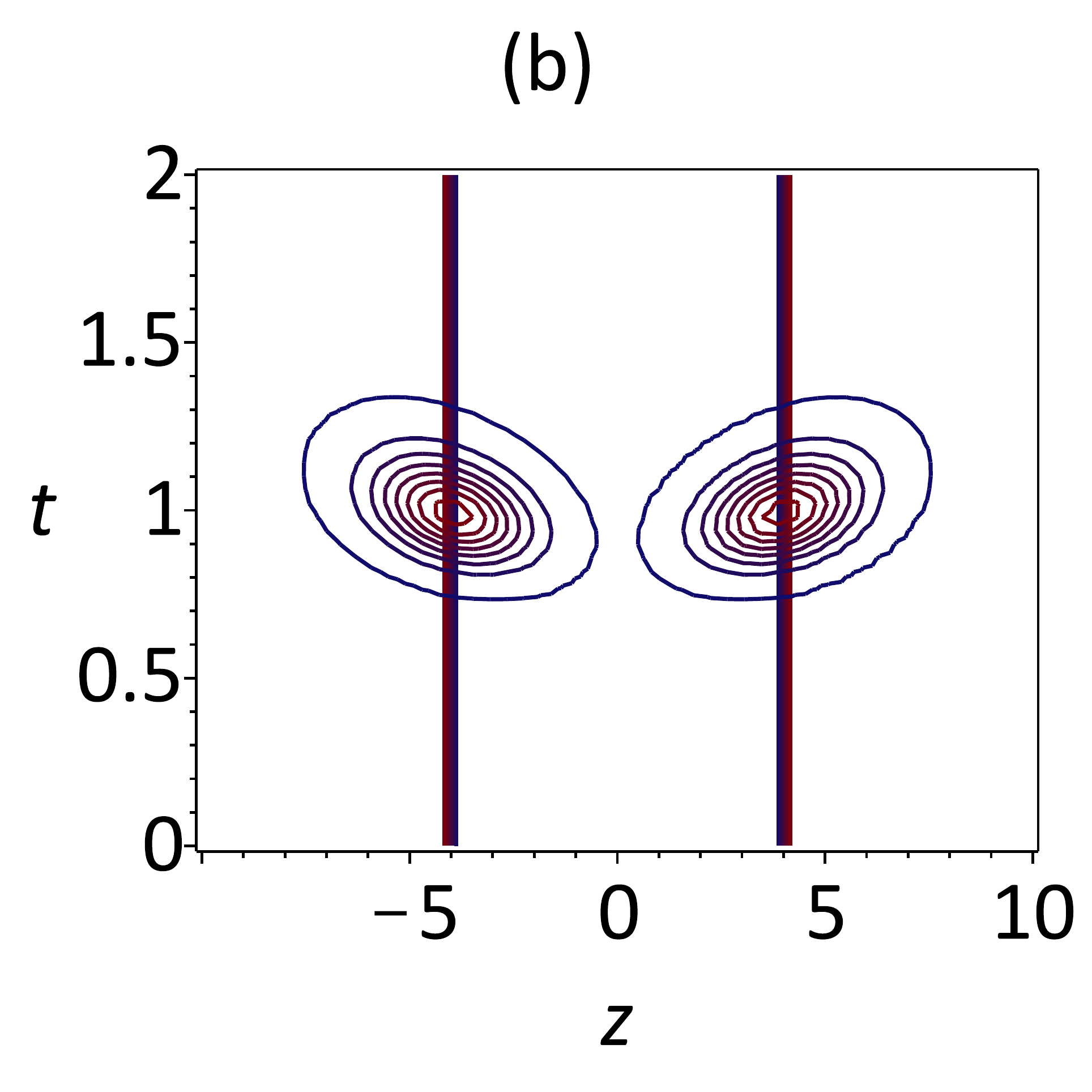}
\bigskip
\par\end{centering}
\par
\bigskip
\par
\bigskip
\par
\bigskip
\caption{The sum of the time dependent densities in the vertical direction
as obtained from addition of Eqs. (\protect\ref{4.23}) and (\protect\ref{4.24}).
Panels a and b show the densities obtained for the weak and strong fields ($%
\protect\alpha =1/2,4$), respectively. The densities are normalized to unit
height. Contours are plotted for the values $%
0.7,0.6,0.5,0.4,0.3,0.2,0.1,0.01 $. Note the overlap of the densities in the
weak field case and their separation when the field is stronger. The
vertical bars denote the location of the maxima of the separate up and down
densities. }
\label{density}
\end{figure}

In Figure \ref{density} we plot the sum of the densities as a function of
the location on the vertical axis $z_{0}$ and time for a weak field (panel
(a), $\alpha =1/2$) and a strong field (panel (b), $\alpha =4$) for $%
y_{s}=-y_{i}=k_{y}=10$. As discussed above, the location of the post-selected
state will be such that the density maximizes about the center of the
Gaussian. Using the notation for the exponent of the spin up spatial density%
\begin{equation}
\rho \left( z,t\right) \equiv \frac{\left( y_{s}-k_{y}t\right) ^{2}}{\left(
1+t^{2}\right) }+\frac{\left( z_{0}-\alpha t\right) ^{2}}{2\left( 1+\frac{%
t^{2}}{4}\right) }  \label{4.25}
\end{equation}%
we readily find that the center of the up density density in space and time
is:%
\begin{equation}
z_{0}=\alpha {\bar t} =\frac{\alpha y_{s}}{k_{y}}.  \label{4.26}
\end{equation}%
Furthermore, in anticipation of a steepest descent estimate of the time
integrals we note that
\begin{equation}
\frac{1}{2}\frac{d^{2}\rho \left( z_{0},t\right) }{dt^{2}}|_{t= {\bar t}}=%
\left[ \frac{k_{y}^{2}+y_{s}^{2}}{\left( 1+{\bar t}^{2}\right) ^{2}}+\frac{\alpha
^{2}}{2\left( 1+\frac{{\bar t}^{2}}{4}\right) }\right]  \label{4.27}
\end{equation}%
This implies that within a steepest descent estimate the normalization
integral is 
\begin{eqnarray}
N\left( \Phi ;\Psi \right) &=&\frac{1}{2}\int_{0}^{\infty }dt\left\vert
\langle \phi _{f},y_{s}|\phi _{+}\rangle \right\vert ^{2}\left[ 1+\exp
\left( -2Y_{t}\right) \right]  \notag \\
&=&\frac{1}{2}\frac{\sqrt{2\pi \left( 1+{\bar t}^{ 2}\right) }\exp \left( -%
\frac{\alpha ^{2}}{2}\right) }{\sqrt{\left( 1+\frac{{\bar t}^{2}}{4}\right) }%
\sqrt{2\left( k_{y}^{2}+y_{s}^{2}\right) \left( 1+\frac{{\bar t}^{2}}{4}%
\right) +\alpha ^{2}\left( 1+{\bar t}^{2}\right) ^{2}}}  \notag \\
&\cdot &\left[ 1+\exp \left( \frac{-4\alpha ^{2}{\bar t}^{2}\left(
k_{y}^{2}+y_{s}^{2}\right) }{\left[ 2\left( k_{y}^{2}+y_{s}^{2}\right)
\left( 1+\frac{{\bar t}^{2}}{4}\right) +\alpha ^{2}\left( 1+{\bar t}^{2}\right)
^{2}\right] }\right) \right]  \label{4.28}
\end{eqnarray}

At this point, we will further specify the pre- and post-selected states.
Following the experiments of Sponar et al \cite{sponar} we choose them to be%
\begin{equation}
\alpha _{i}=\alpha _{f}=\frac{1}{\sqrt{2}},\beta _{i}=\frac{1}{\sqrt{2}}\exp
\left( i\chi _{i}\right) ,\beta _{f}=\frac{1}{\sqrt{2}}\exp \left( i\chi
_{f}\right)  \label{4.29}
\end{equation}%
where the phases $\chi _{i}$ and $\chi _{f}$ are in principle experimentally
controllable phases. The weak values of the spin simplify to%
\begin{eqnarray}
S_{x,w} &=&\frac{\cosh \left( Y_{t}\right) \cos \left( \chi _{f}\right)
+\cos \left( \chi _{i}+\eta _{t}\right) -i\sinh \left( Y_{t}\right) \sin
\left( \chi _{f}\right) }{\cosh \left( Y_{t}\right) +\cos \left( \chi
_{i}-\chi _{f}+\eta _{t}\right) }  \label{4.30} \\
S_{y,w} &=&\frac{\cosh \left( Y_{t}\right) \sin \left( \chi _{f}\right)
+\sin \left( \chi _{i}+\eta _{t}\right) +i\sinh \left( Y_{t}\right) \cos
\left( \chi _{f}\right) }{\cosh \left( Y_{t}\right) +\cos \left( \chi
_{i}-\chi _{f}+\eta _{t}\right) }  \label{4.31} \\
S_{z,w} &=&\frac{\sinh \left( Y_{t}\right) -i\sin \left( \chi _{i}-\chi
_{f}+\eta _{t}\right) }{\cosh \left( Y_{t}\right) +\cos \left( \chi
_{i}-\chi _{f}+\eta _{t}\right) }  \label{4.32}
\end{eqnarray}%
The transition path time distribution is%
\begin{equation}
P_{\Psi }\left( t;\Phi \right) =\frac{\frac{1}{2}\left\vert \langle \phi
_{f},y_{s}|\phi _{+}\rangle \right\vert ^{2}\left[ 1+\exp \left(
-2Y_{t}\right) \right] }{\frac{1}{2}\int_{0}^{\infty }dt\left\vert \langle
\phi _{f},y_{s}|\phi _{+}\rangle \right\vert ^{2}\left[ 1+\exp \left(
-2Y_{t}\right) \right] }\equiv \frac{\left\vert \left\langle \Phi |\Psi
_{t}\right\rangle \right\vert ^{2}}{N\left( \Phi ;\Psi \right) }
\label{4.33}
\end{equation}%
and its steepest descent estimate is:%
\begin{equation}
P_{\Psi }\left( t;\Phi \right) =\frac{1}{\sqrt{\pi }}\sqrt{\left[ \frac{%
k_{y}y_{s}} {{\bar t} \left( 1+{\bar t}^{2}\right) }+\frac{\alpha ^{2}}{2\left(
1+\frac{{\bar t}^{2}}{4}\right) }\right] }\exp \left( -\left[ \frac{k_{y}y_{s}%
} {{\bar t} \left( 1+{\bar t}^{2}\right) }+\frac{\alpha ^{2}}{2\left( 1+\frac{%
{\bar t}^{2}}{4}\right) }\right] \left( t- {\bar t} \right) ^{2}\right) .
\label{4.34}
\end{equation}%
We then have all the information needed to determine the time averaged weak
values and variances and thus to study the weak value uncertainties.

\begin{figure}[tbp]
\begin{centering}
\includegraphics[scale=0.4]{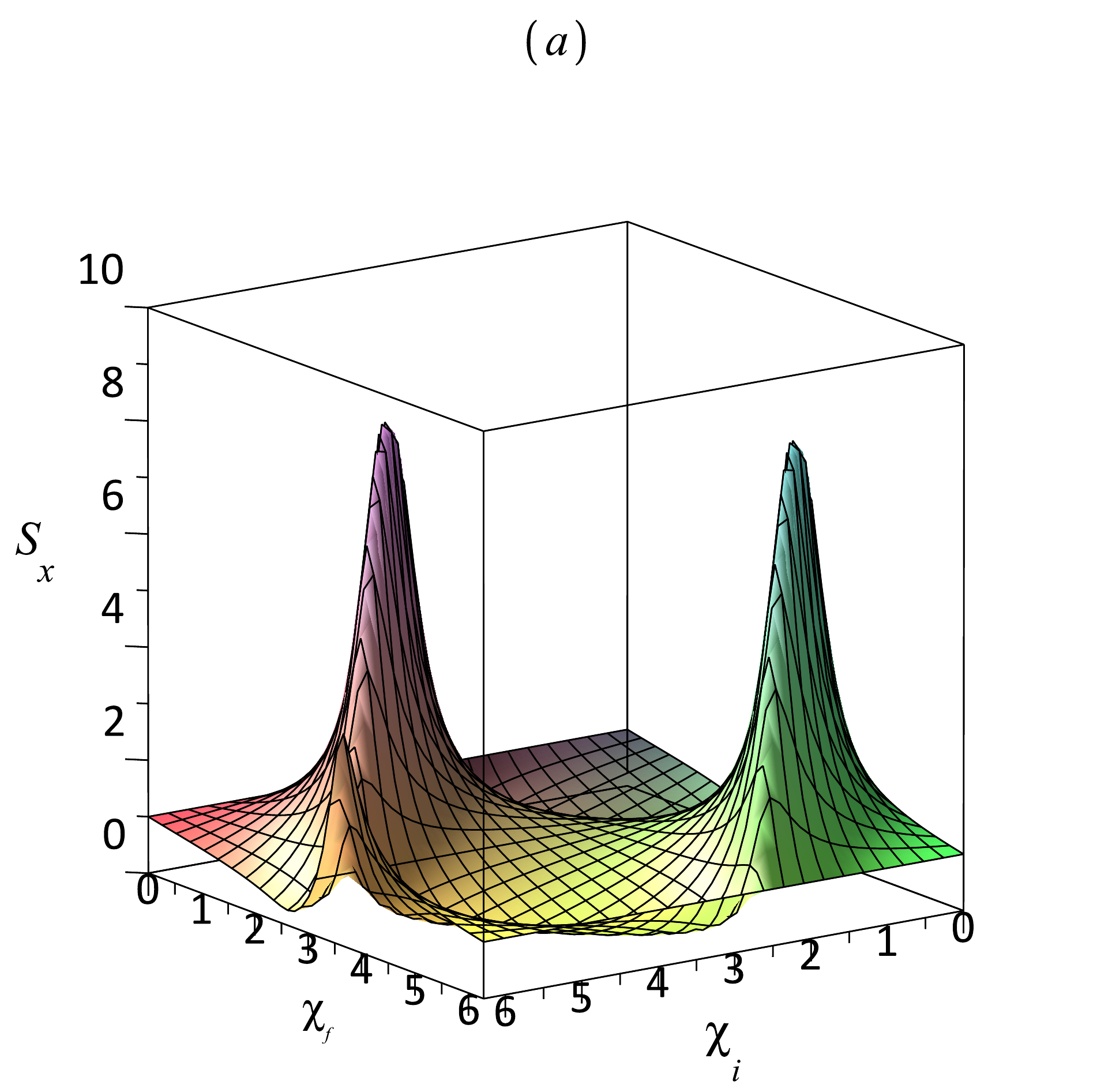}
\includegraphics[scale=0.4]{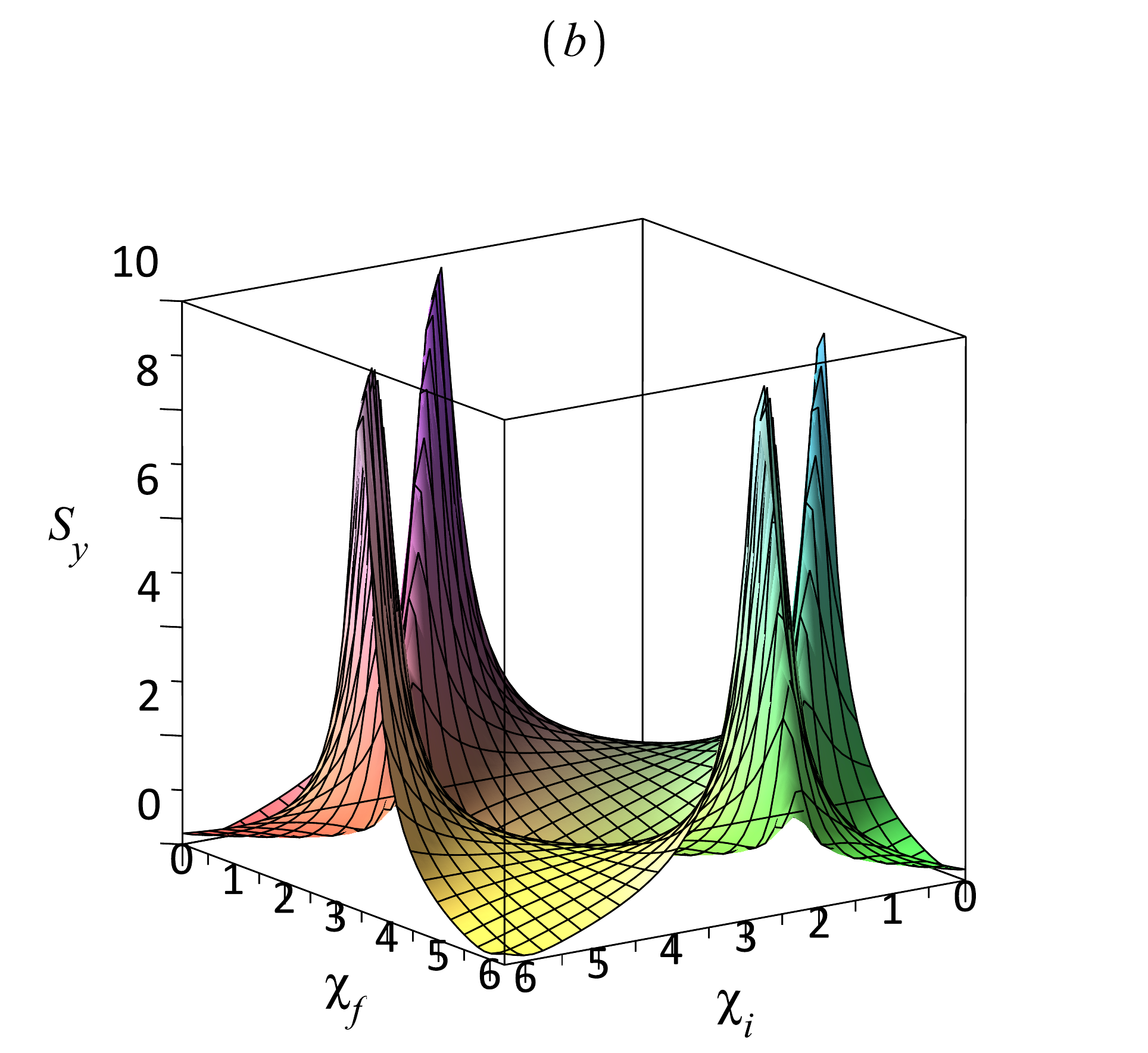}
\bigskip

\bigskip

\bigskip

\bigskip

\includegraphics[scale=0.4]{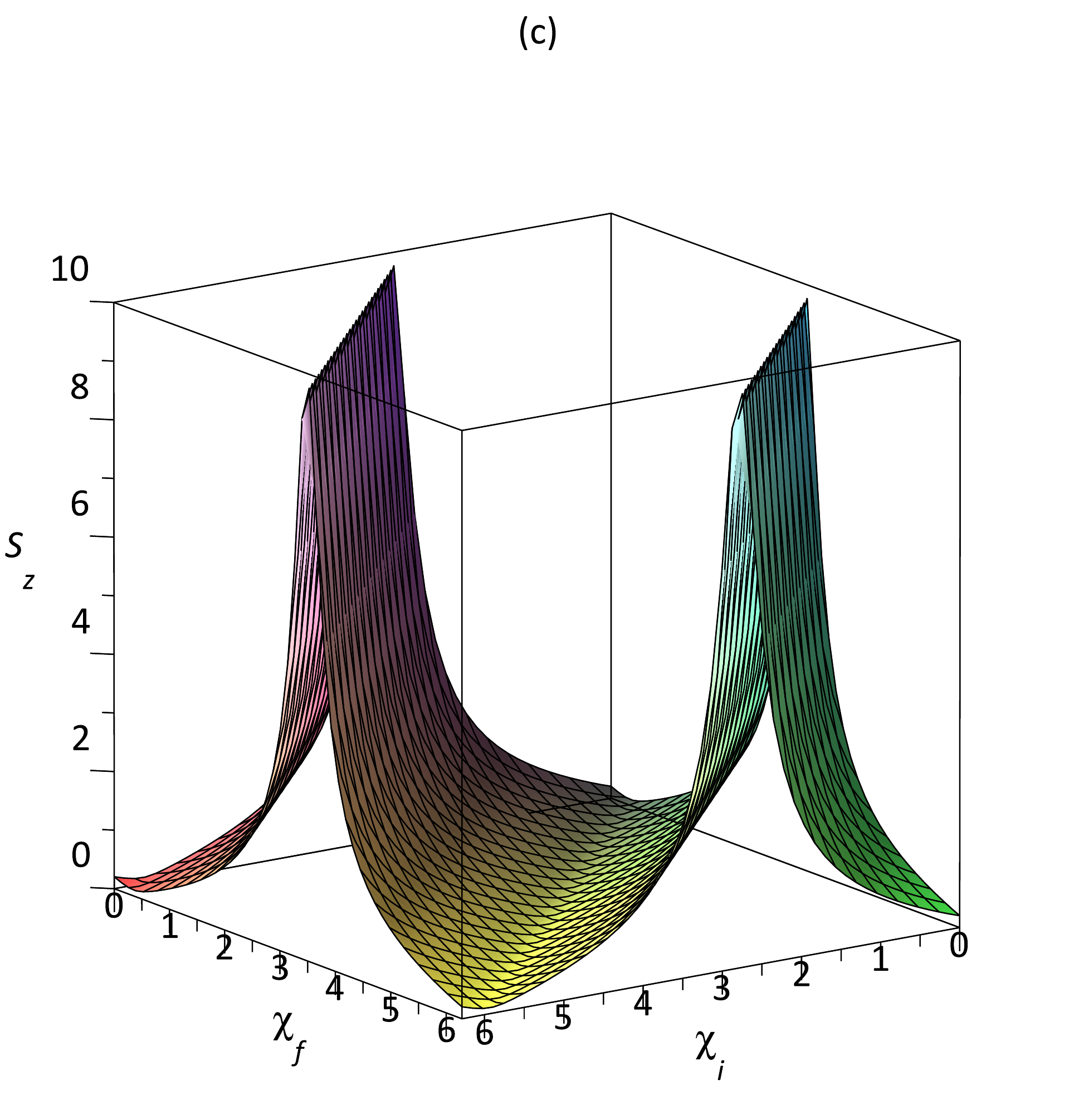}
\bigskip

\par\end{centering}
\par
\bigskip
\par
\bigskip
\par
\bigskip
\caption{The pre- and post-selected phase dependence of the time averaged
absolute values of the spin weak values for a weak field ($\protect\alpha %
=1/2$). Panels a-c show the results for the $x,y$ and $z$ components,
respectively. Note that when the field is weak the absolute magnitude of the
time averaged weak values can become much larger than unity. }
\label{savsmall}
\end{figure}

For implementation we choose $k_{y}=10,y_{s}=10$. This implies that $%
{\bar t}=1$. One should then distinguish between a weak field $\left(
\alpha <1\right) $ and a strong field $\left( \alpha >1\right) $. In the
strong field case $Y_{{\bar t}}\gg 1$ so that in this limit Eqs. (\ref{4.30})-(\ref{4.32}) imply that for all three spin values $%
\left\vert S_{j,w}\right\vert ^{2}\rightarrow 1$ and nothing interesting
occurs. The situation differs drastically when the field is weak. In panels
a-c of Fig. \ref{savsmall} we plot the absolute value of the time averaged
weak values of the spin in the $x,y$ and $z$ directions for $\alpha =1/2$ as
a function of the pre- and post-selected phases $\chi _{i}$ and $\chi _{f}$.
One notices that  when the field is weak one obtains substantial
deviations from unity.

\begin{figure}[tbp]
\begin{centering}
\includegraphics[scale=0.4]{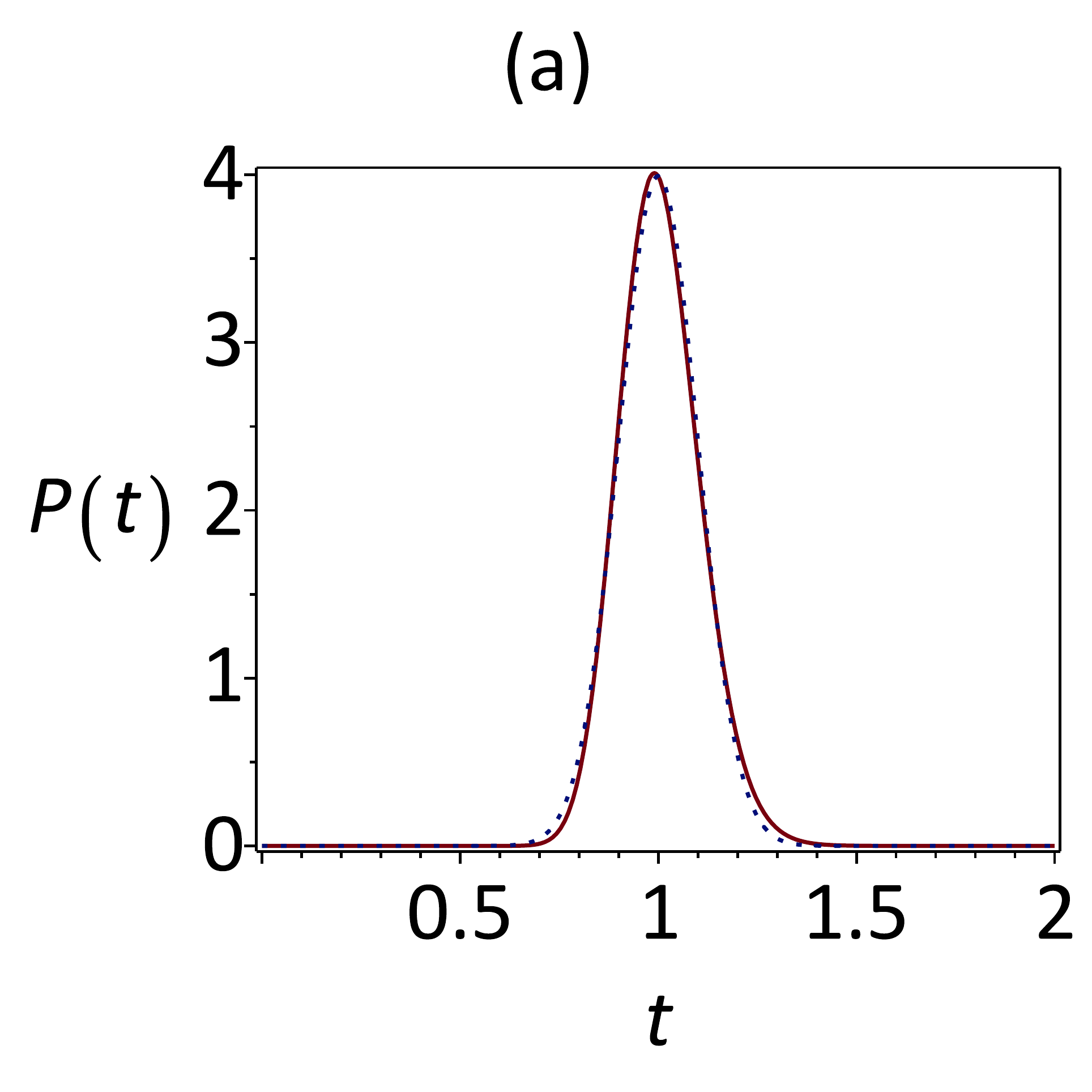}
\includegraphics[scale=0.4]{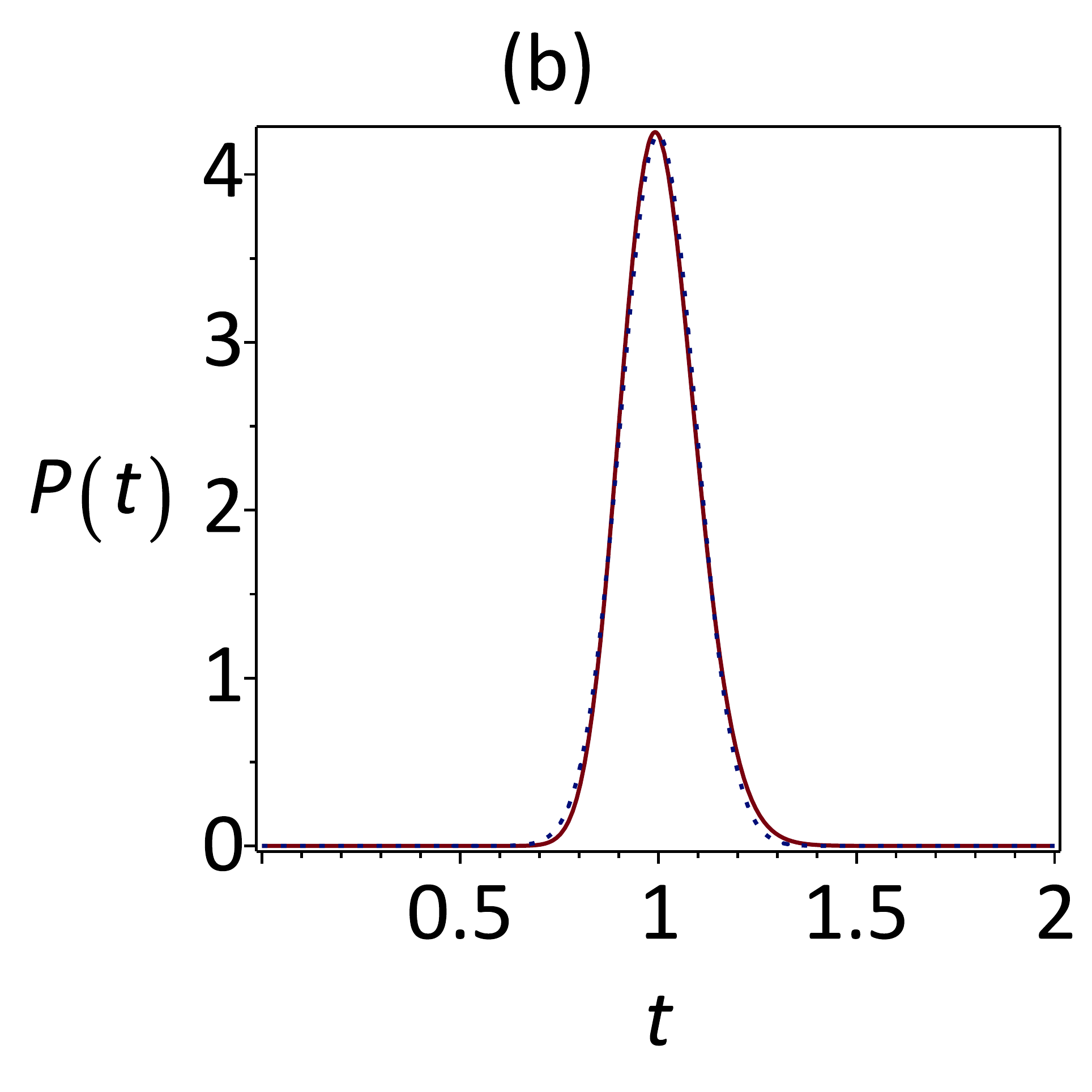}

  \bigskip
\par\end{centering}
\par
\bigskip
\par
\bigskip
\par
\bigskip
\caption{Transition path time distribution for a weak magnetic field ($%
\protect\alpha=1/2$, panel a) and a strong one ($\protect\alpha=4$, panel
b). The solid and dotted lines in each panel are respectively the
numerically exact transition path time distribution and its steepest descent
estimate (Eq. \protect\ref{3.51}). Note the accuracy of the steepest descent
approximation. }
\label{pt}
\end{figure}

The difference between the weak and strong field cases lies in the weak
values, not in the time averaging. The transition path time distribution is
insensitive to the field strength, it mainly reflects the distribution of
momenta in the pre-selected state, as may be seen from Fig. \ref{pt}. It is
very narrow and localized about the time one would expect from the motion of
a free particle whose momentum is $10$ and the distance travelled is $10$.
Thus the differences between the weak and strong fields for the weak value
spins reflects the central difference between the effect of strong and weak
fields on the spin weak values. In a weak field, the overlap of pre- and
post-selected state is a sensitive function of the phases, it is not so for
the strong field. This of course means that only when using a weak field
will one be able to discern the phases from the weak values. They are
unimportant in the strong field case.

The time averaged values of the spins may be estimated using the
steepest descent approximation. The leading order contribution is 
\begin{equation}
\left\langle S_{j,w}\left( t\right) \right\rangle \simeq S_{j,w}\left(
{\bar t}\right) +\frac{1}{4}\frac{d^{2}S_{j,w}\left( t\right) }{dt^{2}}%
|_{t={\bar t}}\left[ \frac{k_{y}y_{s}}{{\bar t} \left( 1+{\bar t}^{ 2}\right) }%
+\frac{\alpha ^{2}}{2\left( 1+\frac{{\bar t}^{2}}{4}\right) }\right] ^{-1}.
\label{4.35}
\end{equation}%
The variance of the time averaged weak spin values may be approximated as:
\begin{eqnarray}
\left\langle S_{j,w}\left( t\right) S_{j,w}^{\ast }\left( t\right)
\right\rangle -\left\vert \left\langle S_{j,w}\left( t\right) \right\rangle
\right\vert ^{2} &\simeq &\left\vert \frac{dS_{j,w}\left( t\right) }{dt}%
\right\vert _{t={\bar t}}^{2}\left\langle \left( t-{\bar t}\right)
^{2}\right\rangle  \notag \\
&=&\frac{1}{2}\left\vert \frac{dS_{j,w}\left( t\right) }{dt}\right\vert
_{t={\bar t}}^{2}\left[ \frac{k_{y}y_{s}}{{\bar t}
\left( 1+{\bar t}^{2}\right) }+\frac{\alpha ^{2}}{2\left( 1+\frac{{\bar t}^{2}}{4}\right) }\right]
^{-1}.  \label{4.36}
\end{eqnarray}

\begin{figure}[tbp]
\begin{centering}
\includegraphics[scale=0.4]{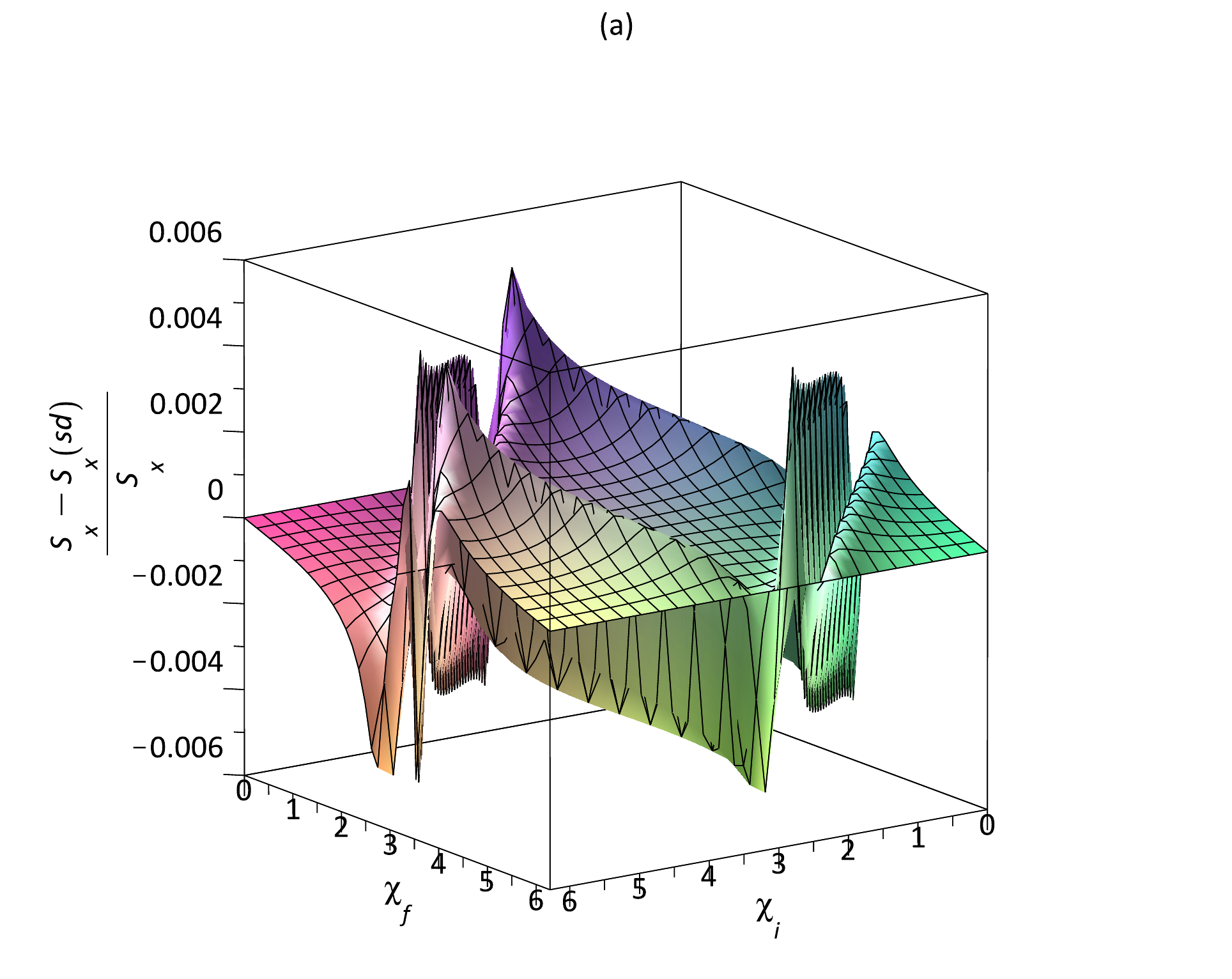}
\includegraphics[scale=0.4]{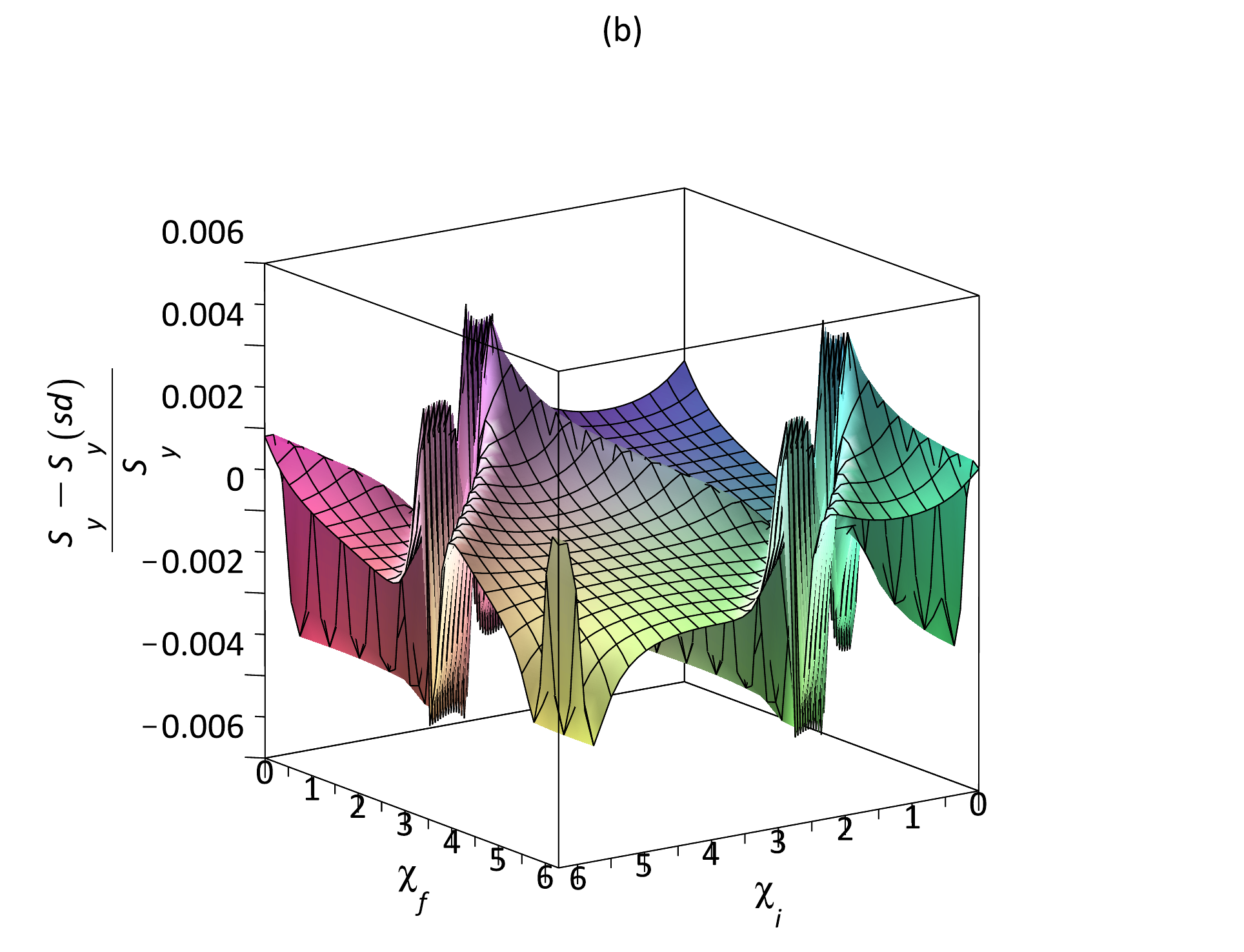}
\includegraphics[scale=0.4]{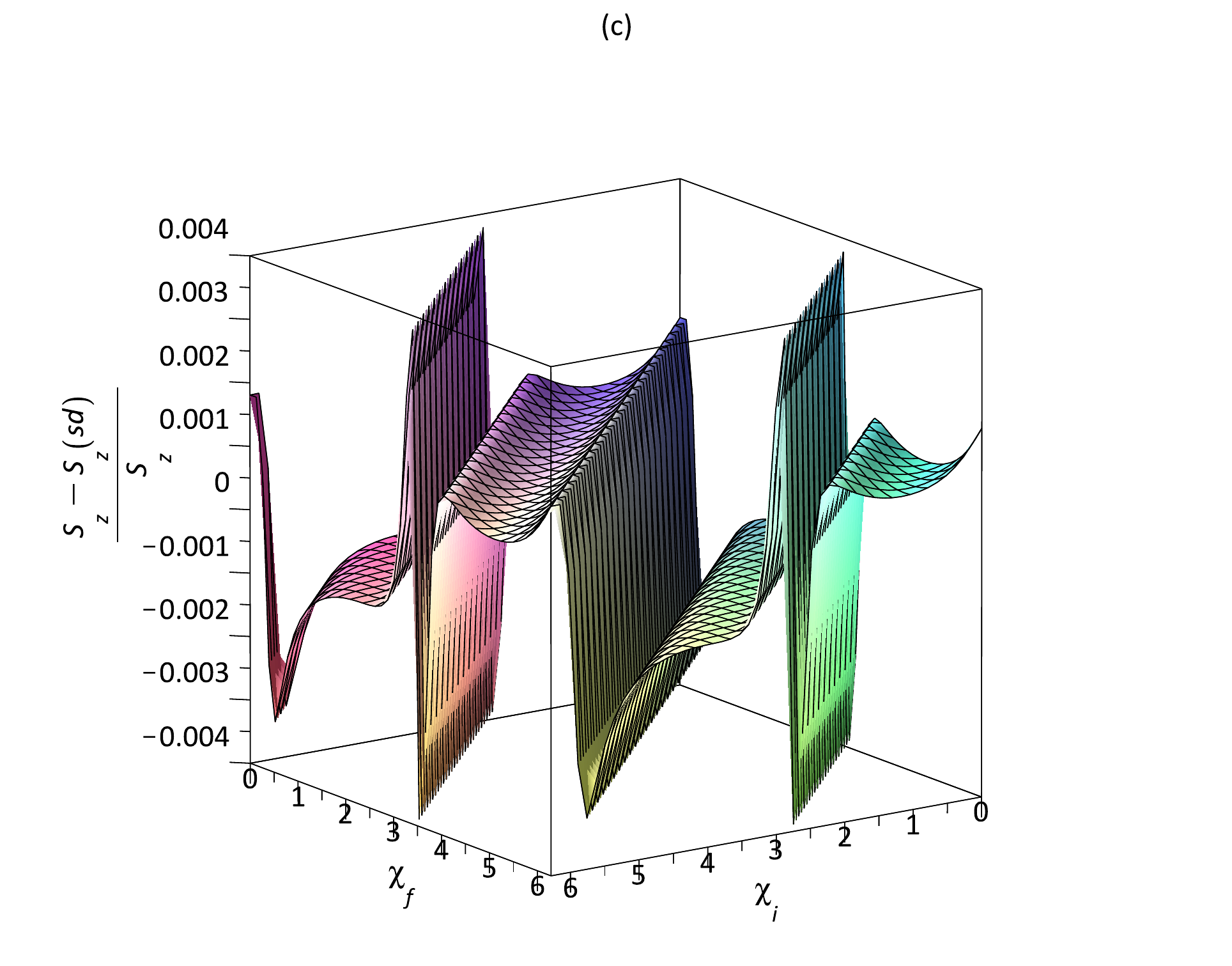}
\bigskip
\par\end{centering}
\par
\bigskip
\par
\bigskip
\par
\bigskip
\caption{The accuracy of the leading order steepest descent estimate of the
time averaged mean value of the absolute values of the spins in the $x,y$
and $z$ directions (panels a-c) in the low field case ($\protect\alpha=1/2$%
). Note that in all cases, the error is less than one percent. }
\label{sderror}
\end{figure}

To make further progress we check the accuracy of the steepest descent
approximation to the time averaged weak values of the absolute value of the
three spins. In Fig. \ref{pt} we show that the steepest descent
approximation to the transition path time distribution is quite accurate.
Then in Fig. \ref{sderror} we plot the relative error for the time averaged
spin weak values obtained using the leading order term only for the steepest
descent time average of the spins, $S_{j,w}({\bar t})$, as compared to the
numerically exact results. The agreement is excellent, the error is in all
cases $\sim 0.5\%$ or lower.

\begin{figure}[tbp]
\begin{centering}
\includegraphics[scale=0.4]{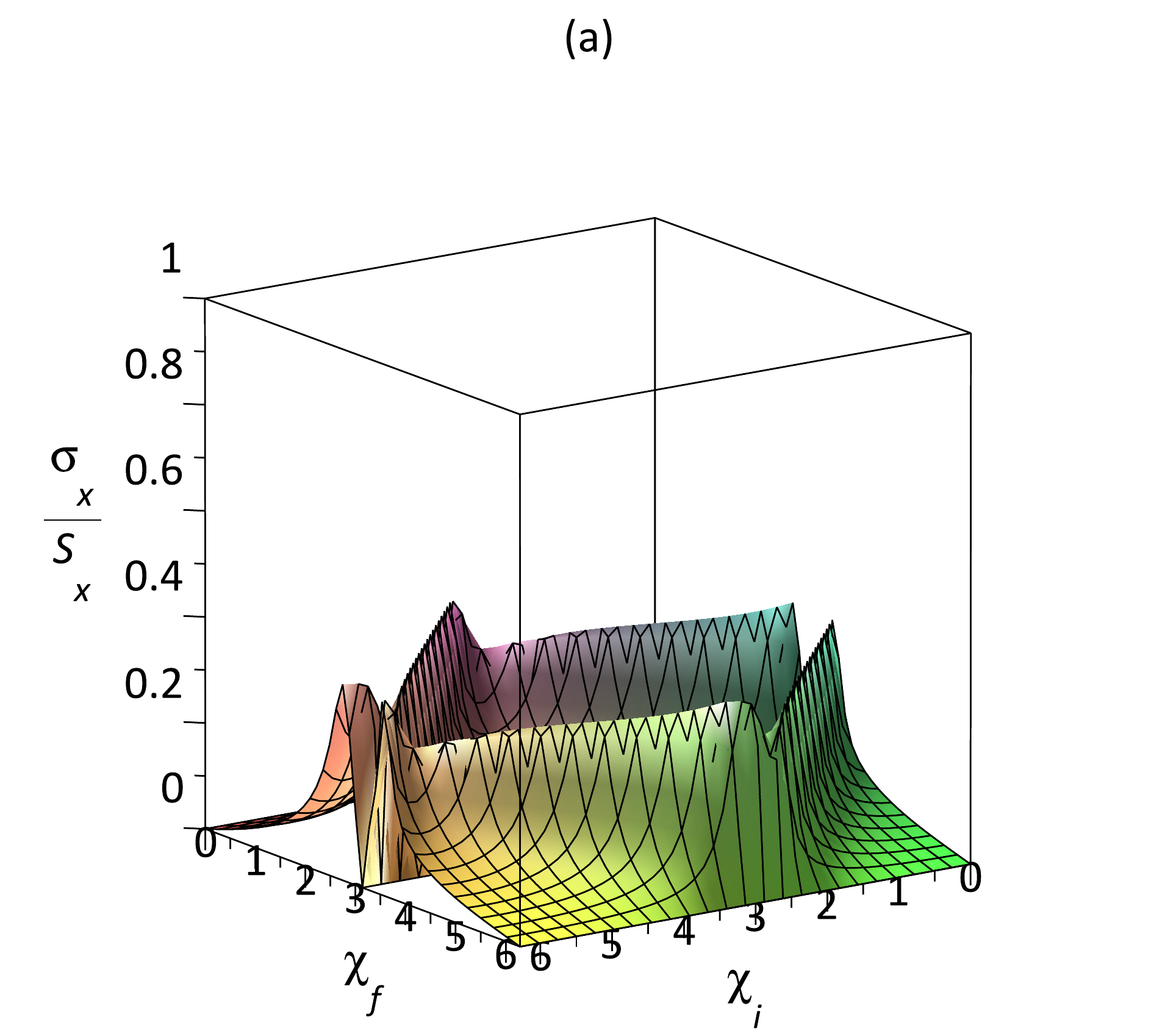}
\includegraphics[scale=0.4]{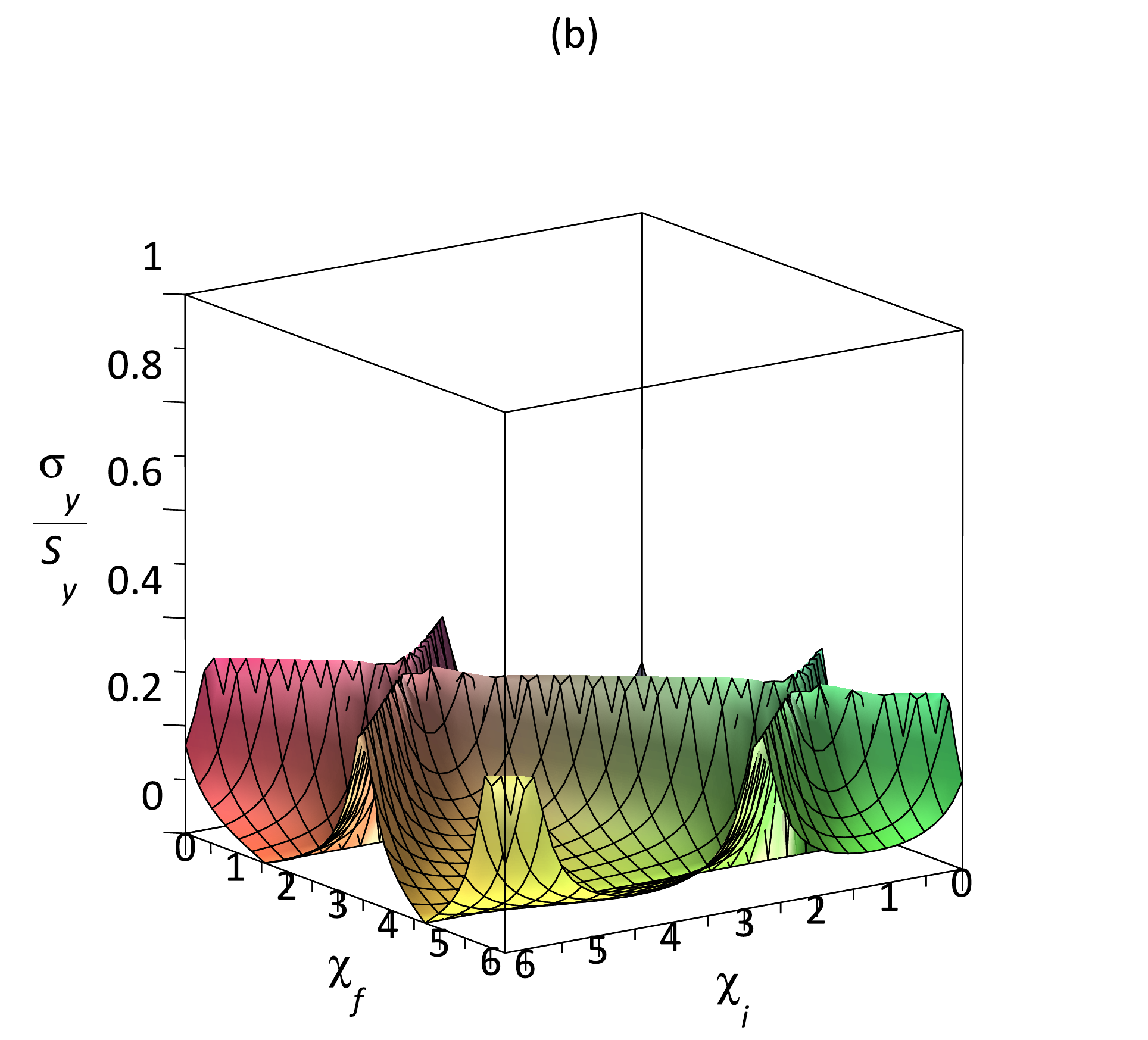}
\includegraphics[scale=0.4]{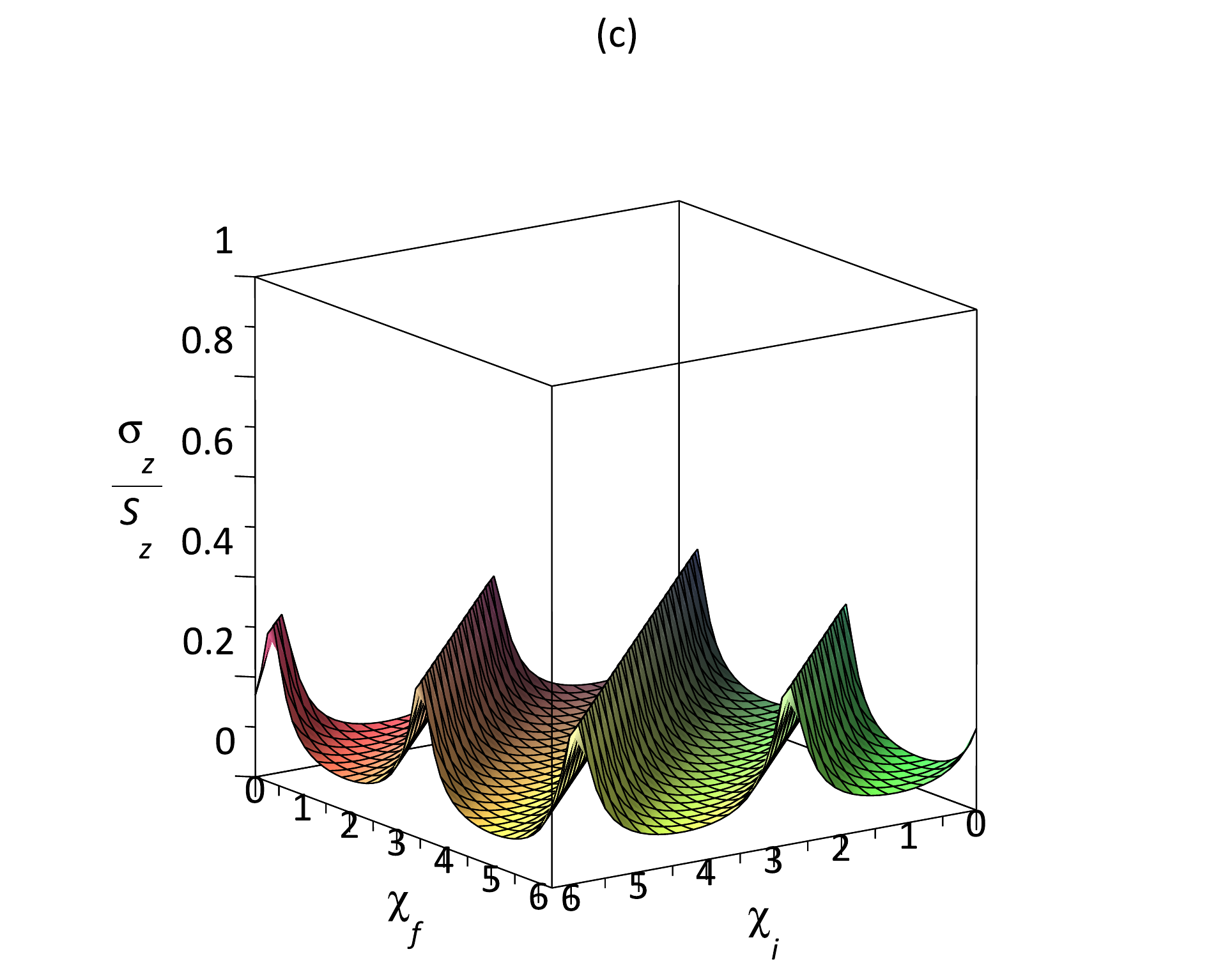}  \bigskip
\par\end{centering}
\par
\bigskip
\par
\bigskip
\par
\bigskip
\caption{The ratio of the standard deviation of the time averaged spin weak
values for a strong field to the magnitude of the spin weak value. Panels
a-c show the results for the $x,y$ and $z$ directions, respectively. }
\label{sigmas}
\end{figure}

This then enables us to obtain the standard deviation for the time averaged
weak values, defined as $\sigma _{j}=\sqrt{\langle \Delta S_{j,w}^{2}\rangle
}$ obtained from the steepest descent estimate as in Eq. (\ref{3.52}). The
results for the weak field ($\alpha =1/2$) case are shown in Fig. \ref%
{sdevlow}.
\begin{figure}[tbp]
\begin{centering}
\includegraphics[scale=0.4]{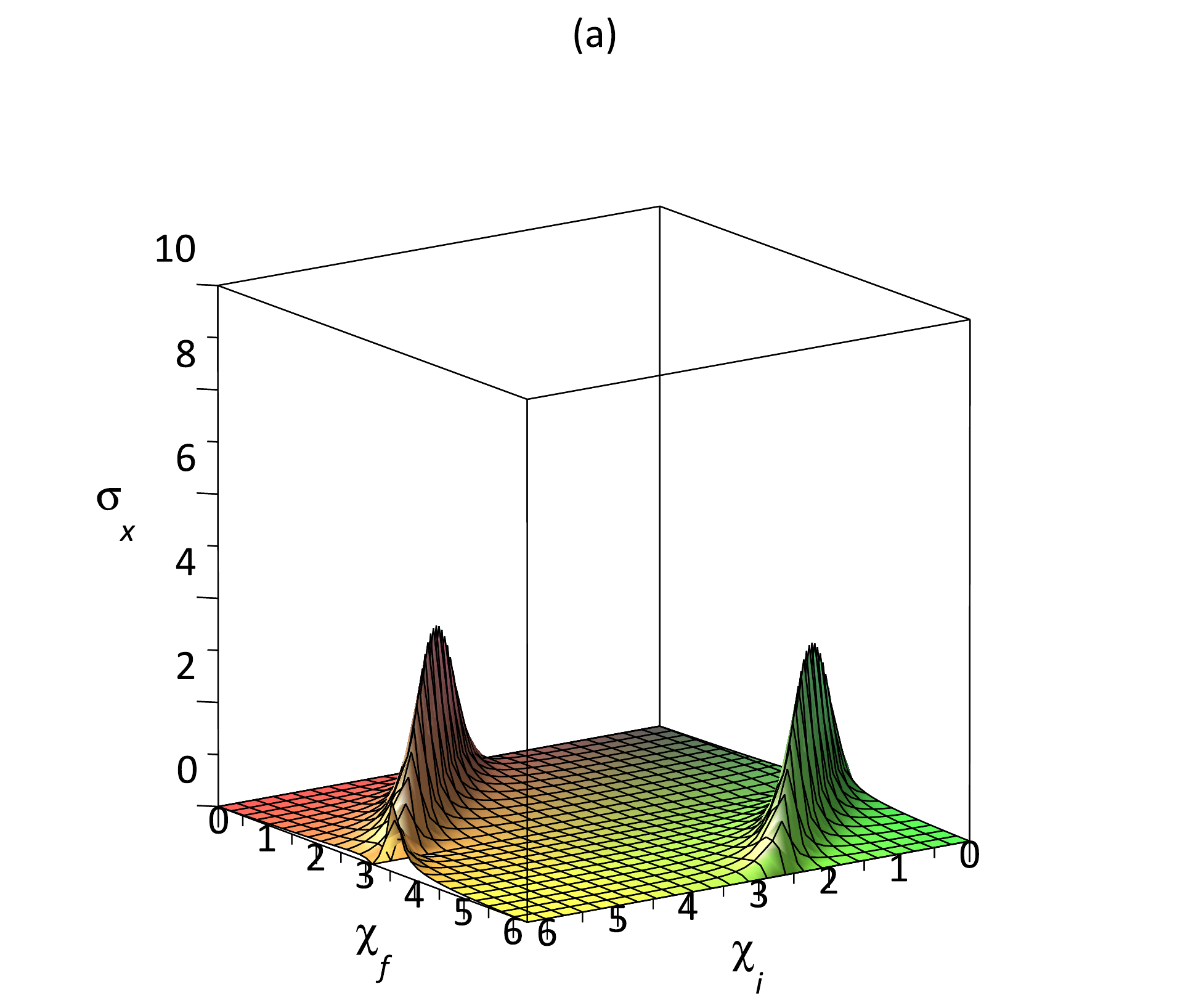}
\includegraphics[scale=0.4]{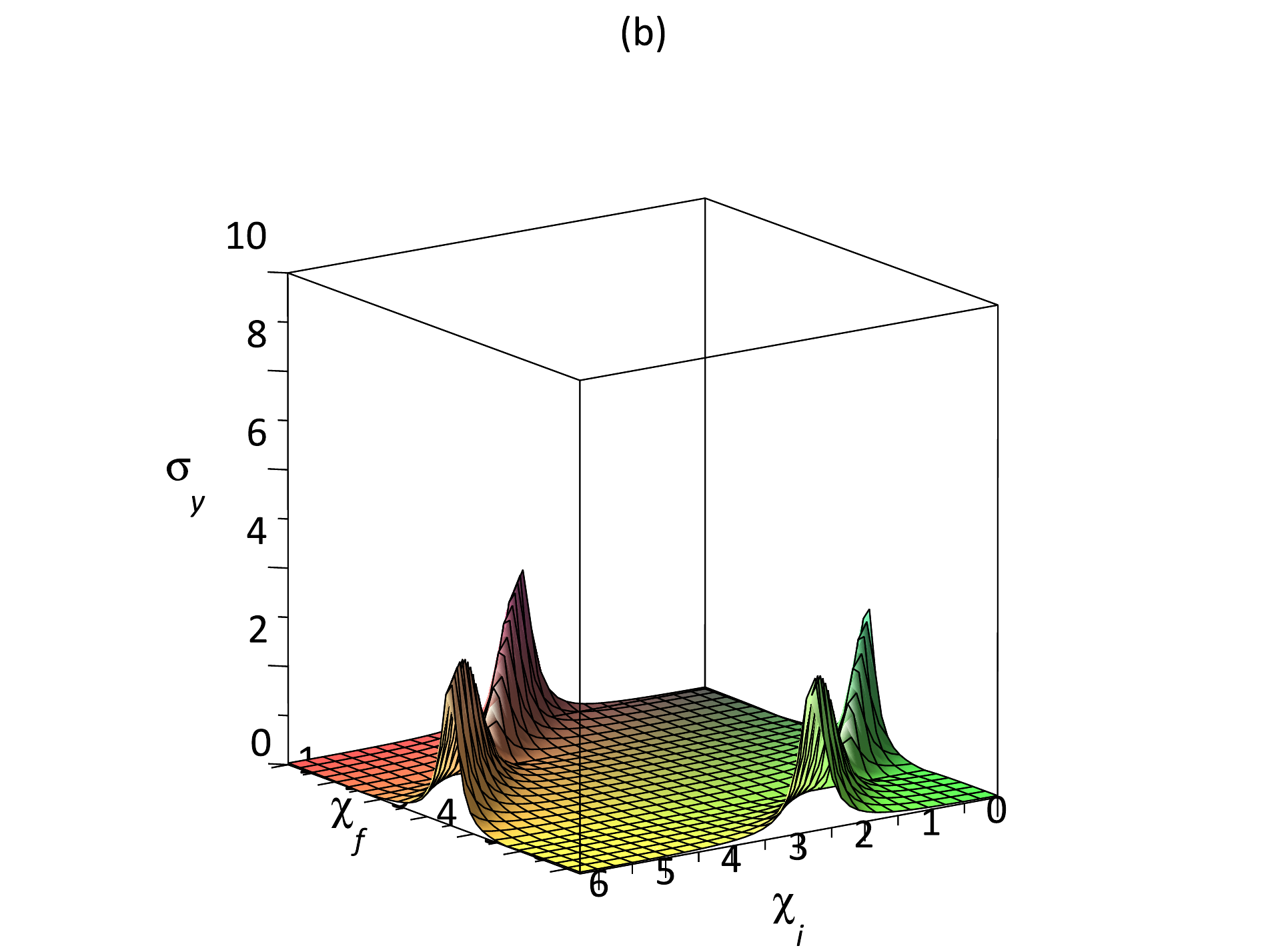}
\includegraphics[scale=0.4]{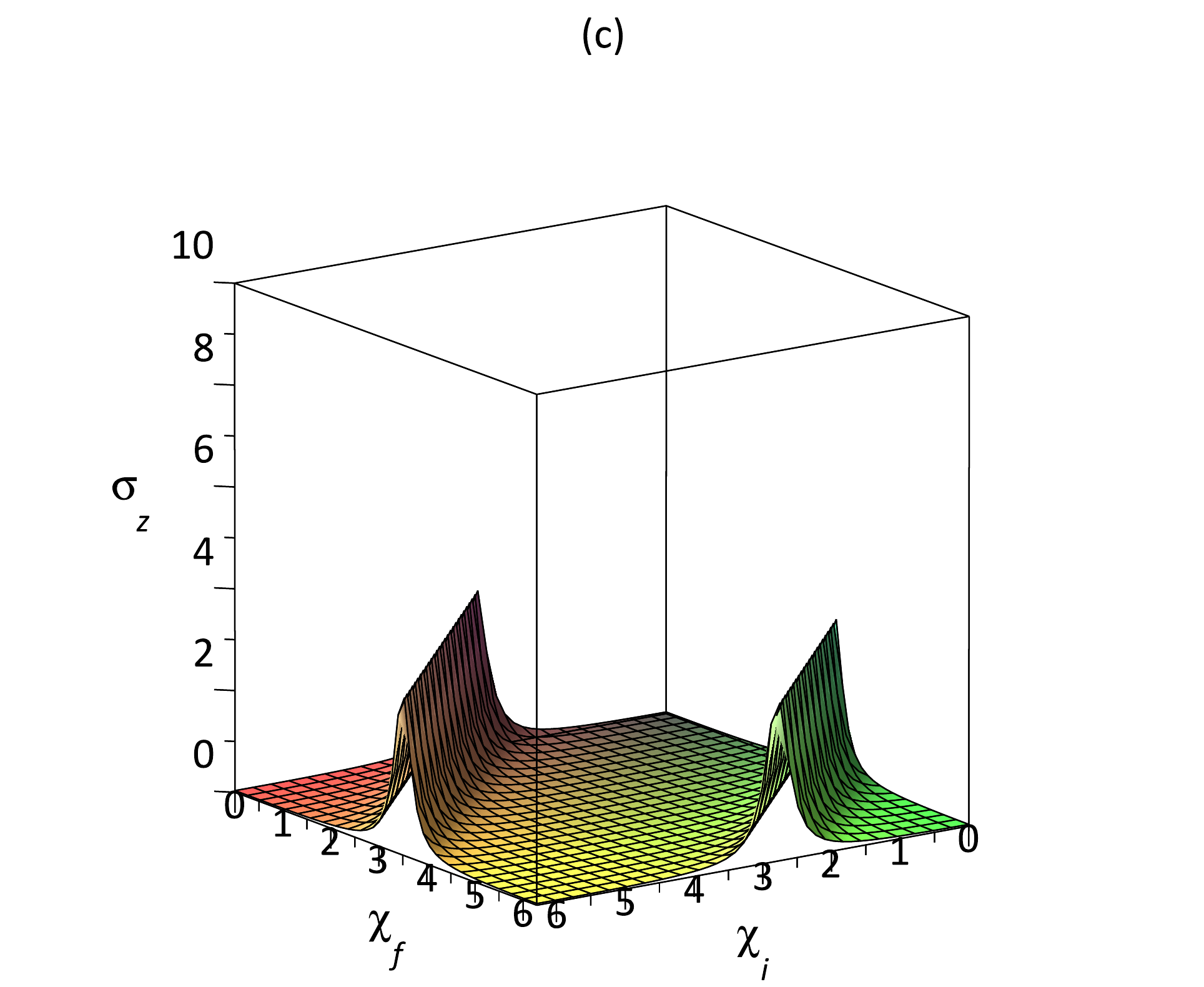}  \bigskip
\par\end{centering}
\par
\bigskip
\par
\bigskip
\par
\bigskip
\caption{The standard deviation of the time averaged spin weak values for a
weak field. Panels a-c show the results for $\sigma_{x},\sigma_{y}$ and $\sigma_{z}$
respectively for a field strength of $\protect\alpha =1/2$. }
\label{sdevlow}
\end{figure}
As anticipated from the general weak value uncertainty principle, we find
that when the weak values are much larger than unity in magnitude there also
is a relatively large standard deviation of $\sim 30\%$ confirming that for
anomalous weak values one needs more averaging of the experimental signal to determine them. This is exemplified when considering the ratio of the standard deviation to the
magnitude of the spin as shown in the weak field ($\alpha =1/2$) case in Fig. %
\ref{sigmas}.

Some further comments about the time averaged weak values are in order. Denoting
\begin{equation*}
w=\chi _{i}-\chi _{f}
\end{equation*}%
one finds from Eq. (\ref{4.32}) that within the steepest descent (sd) approximation%
\begin{equation*}
\left\vert \left\langle S_{z,w}\right\rangle _{sd}\right\vert ^{2}=\frac{%
\cosh \left( Y_{_{{\bar t}}}\right) -\cos \left( w+\eta _{{\bar t}}\right) }{\cosh
\left( Y_{_{{\bar t}}}\right) +\cos \left( w+\eta _{_{{\bar t}}}\right) }.
\end{equation*}%
The value of $w$ that maximizes the weak value in the vertical direction is
\begin{equation*}
w+\eta _{t}=\left( 2m+1\right) \pi
\end{equation*}%
so that
\begin{equation*}
\max \left\vert \left\langle S_{z,w}\right\rangle _{sd}\right\vert ^{2}=%
\frac{\cosh \left( Y_{_{{\bar t}}}\right) +1}{\cosh \left( Y_{_{{\bar t}}}\right) -1}=\coth ^{2}\left( \frac{Y_{_{{\bar t}}}}{2}\right) .
\end{equation*}%
As the field strength $\alpha $ grows weaker, $Y_{_{{\bar t}}}$ becomes
smaller and the weak value grows, that is why when $\alpha =1/2$ we find the
maximal value of $\simeq 10$ for the magnitude of the weak value.

\section{Discussion and conclusion}

A weak value uncertainty relation was derived for two not necessarily Hermitian operators. For this purpose, weak value operators were defined and their basic algebra presented. The uncertainty relation was obtained by noting that in principle, weak values may be time dependent. The time averaged means and variances of the weak values were obtained through a suitably defined transition
path time distribution. The weak value uncertainty relation relates the variances of the weak value operators to the mean values of their commutators and anti-commutators.  The resulting weak value uncertainty relation can be considered as a direct analog of the strong value (Robertson) uncertainty relation.  An important consequence of the weak
value uncertainty relation is that it is possible to know, with high accuracy, the simultaneous mean weak values of non-commuting operators by employing a judicious
choice  of the pre- and post-selected states.

To illustrate the utility of the uncertainty relation several examples have been analyzed.
First, the weak value time-energy uncertainty relation has been evaluated and reduced to the standard value of $\hbar^2 / 4$ when initially
the overlap of the pre- and post-selected states vanishes. Second, due to the proportionality found between the weak values of the momentum
and coordinate when the post-selected state is a coherent state, both quantities can be known simultaneously with arbitrary precision even though the
two operators themselves do not commute.  Third, the weak value coordinate-kinetic energy uncertainty relation when the post-selected state is also a
coherent state shows that uncertainty in the coordinate also leads to an uncertainty in the kinetic energy. And finally, the same analysis has
been carried out for cyclic operators such as the spin operators when analyzing the Stern-Gerlach experiment. The
fluctuations of the weak values of the spin also display a proportionality implying that the uncertainty principle does not impose any restriction
on the spin weak values even if their corresponding operators do not commute. The weak value uncertainty relation leads to the observation that the variance of the spin weak value is proportional to its magnitude. This means that
for anomalously large weak values, the variance is also large, so that from an experimental point of view, more measurements on the
system have to be carried out in order to obtain the precise mean anomalously large weak value as compared to the "standard" spin values.

The time averaging of weak values which lies at the heart of the weak value uncertainty relation has not been considered previously. As laid out in the original paper by Aharonov and coworkers \cite{aharonov88}, the weak measurement of an operator is implemented by coupling the system to the measuring device through the same operator. As a result, the time evolution operator commutes with the operator whose weak value is measured so that the weak value becomes time independent. Under such conditions, the time averaging considered in the present paper becomes trivial and unnecessary. However, our analysis of the Stern-Gerlach experiment,  based on the scenario presented in Ref. \cite{aharonov88} shows that in actuality one cannot isolate the spin component from the spatial component so that the time evolution becomes non-trivial and impacts the measured weak values. In this sense, one should consider the original scenario suggested in Ref. \cite{aharonov88} as expressing the time averaged spin weak values. As shown, these too can be anomalously large, verifying the results of Aharonov et al.


\vspace{2cm}

\noindent
{\bf Acknowledgement}
We thank E. Cohen and L. Vaidman for illuminating discussions. This work was supported by a grant from the Israel Science Foundation and was partially supported by a grant with Ref.
FIS2017-83473-C2-1-P from the Ministerio de Ciencia, Innovaci\'on y Universidades
(Spain).


\vspace{2cm}

\begin{thebibliography}{199}

\bibitem{robertson}
H. P. Robertson, Phys. Rev. {\bf 34}, 163 (1929).

\bibitem{ozawa03} M. Ozawa, Phys. Rev. A \textbf{67}, 042105 (2003).

\bibitem{aharonov88}
Y. Aharonov, D. Z. Albert and L. Vaidman, Phys. Rev. Lett. \textbf{60},  1351 (1988).

\bibitem{wiseman07}
H. M. Wiseman, New J. Phys. \textbf{9}, 165 (2007).

\bibitem{nori12}
A. G. Kofman, S. Ashhab and F. Nori,    Phys. Rep. {\bf 520}, 43 (2012)


\bibitem{hiley14}
R. Flack and B. J. Hiley, J. Phys. Conf. Ser. \textbf{504}, 012016 (2014).

\bibitem{steinberg11}
S. Kocis, B. Braverman, S. Ravets, M. J. Stevens, R. P. Mirin, L. K. Shalm and M. A. Steinberg, Science \textbf{332}, 1170 (2011).


\bibitem{lundeen11}
J. S. Lundeen, B. Sutherland, A. Patel, C. Stewart, C. Bamber, Nature {\bf 474}, 188 (2011).

\bibitem{tamir13}
B. Tamir and E. Cohen, Quanta {\bf 2}, 7 (2013).

\bibitem{dressel14} J. Dressel, M. Malik, F.M. Miatto, A.N. Jordan and R.W. Boyd, Rev. Mod. Phys. {\bf 86}, 307, (2014).

\bibitem{lund10} A. P. Lund and H. M. Wiseman, New J. Phys. \textbf{12}, 093011 (2010).

\bibitem{rozema12} L. A. Rozema, A. Darabi, D. H. Mahler, A. Hayat, Y. Soudagar  and
A. M. Steinberg, Phys. Rev. Lett. \textbf{109}, 100404 (2012).


\bibitem{eli1}
J. Petersen and E. Pollak, J. Phys. Chem. Lett. {\bf 8}, 4017 (2017).

\bibitem{eli2}
E. Pollak and S. Miret-Art\'es, New. J. Phys. {\bf 20}, 073016 (2018).

\bibitem{aharonov61}
Y. Aharonov and D. Bohm, Phys. Rev. {\bf 122}, 1649 (1961).

\bibitem{busch08}
P. Busch, Lect. Notes in Phys. \textbf{734}, 73  (2008).

\bibitem{denkmayr17} T. Denkmayr, H. Geppert, H. Lemmel, M. Waegell, J. Dressel, Y. Hasegawa, and S. Sponar, Phys. Rev. Lett. \textbf{118}, 010402 (2017).

\bibitem{sponar18} S. Sponar, T. Denkmayr, H. Geppert-Kleinrath, Y.
Hasegawa, and J. Dressel, Physica B, in press, DOI:
10.1016/j.physb.2018.04.014 (2018).

\bibitem{cohen18} E. Cohen and E. Pollak, preprint, arXiv:1804.11298 [quant-ph].

\bibitem{cohen-tannoudji}
C. Cohen-Tannoudji, B. Diu  and F. Laloe, Quantum Mechanics, Wiley and Sons, New York, Vol. 1, pp 286-287 (1977).



\bibitem{benitez}
E. Ben\'{i}tez Rodr\'{i}guez, L. M. Ar\'evalo Aguilar and E. Piceno Mart\'{i}nez, Eur. J. Phys. {\bf 38}, 069501 (2017).

\bibitem{sponar}
S. Sponar, T. Denkmayr, H. Geppert, H. Lemmel, A. Matzkin, J. Tollaksen and Y. Hasegawa,
Phys. Rev. A {\bf 92}, 062121 (2015).

\end{thebibliography}
\end{document}